\documentclass[preprint]{emulateapj}

\submitted{Accepted in ApJ}

\usepackage{graphicx}        
\usepackage{bm}              
\usepackage{natbib}
\usepackage[mediumspace,Gray,squaren]{SIunits}

\newcommand{\arone}{148\,GHz}
\newcommand{\artwo}{218\,GHz}
\newcommand{\arthree}{277\,GHz}

\newcommand{\commentx}[1]{}

\renewcommand{\vec}[1]{\mbox{\boldmath$#1$}} 

\newcommand{\ra}[3]   
   {\makebox[1.5em][r]{#1}\makebox[1.5em][r]{#2} \makebox[2em][r]{#3}}

\newcommand{\hms}[3]  
   {${#1}^{\mathrm{h}}{#2}^{\mathrm{m}}{#3}^{\mathrm{s}}$}

\newcommand{\hmin}[2]  
   {\ensuremath{{#1}^{\mathrm{h}}{#2}^{\mathrm{m}}}}
   
\newcommand{\hours}[1]  
   {\ensuremath{{#1}^{\mathrm{h}}}}

\newcommand{\dms}[3]  
   {\ensuremath{{#1}\degree{#2}\arcminute{#3}\arcsecond}}

\newcommand{\dm}[2]  
   {\ensuremath{{#1}\degree{#2}\arcminute}}

\newcommand{\ukcmb}  
           {\ensuremath{\micro \kelvin_\mathrm{cmb}}}

\newcommand{\uk}  
           {\ensuremath{\micro \kelvin}}

\newcommand{\fdeg} 
           {\hbox{$.\!\!^{\circ}$}}

\usepackage{color}

\newcommand{\area}{455 square-degrees}
\newcommand{\significance}{5.25}

\newcommand{\numberOfSources}{157}
\newcommand{\minimumFlux}{15\,mJy}
\newcommand{\maximumFlux}{1500\,mJy}

\begin{document}

\title{The Atacama Cosmology Telescope: Extragalactic Sources at \arone\ in the 2008 Survey}
\author{
T.~A.~Marriage\altaffilmark{1,2},
J.~B.~Juin\altaffilmark{3},
Y.-T.~Lin\altaffilmark{4,1,3},
D.~Marsden\altaffilmark{5},
M.~R.~Nolta\altaffilmark{6},
B.~Partridge\altaffilmark{7},
P.~A.~R.~Ade\altaffilmark{8},
P.~Aguirre\altaffilmark{3},
M.~Amiri\altaffilmark{9},
J.~W.~Appel\altaffilmark{10},
L.~F.~Barrientos\altaffilmark{3},
E.~S.~Battistelli\altaffilmark{11,9},
J.~R.~Bond\altaffilmark{6},
B.~Brown\altaffilmark{12},
B.~Burger\altaffilmark{9},
J.~Chervenak\altaffilmark{13},
S.~Das\altaffilmark{14,10,1},
M.~J.~Devlin\altaffilmark{5},
S.~R.~Dicker\altaffilmark{5},
W.~B.~Doriese\altaffilmark{15},
J.~Dunkley\altaffilmark{16,10,1},
R.~D\"{u}nner\altaffilmark{3},
T.~Essinger-Hileman\altaffilmark{10},
R.~P.~Fisher\altaffilmark{10},
J.~W.~Fowler\altaffilmark{10},
A.~Hajian\altaffilmark{6,1,10},
M.~Halpern\altaffilmark{9},
M.~Hasselfield\altaffilmark{9},
C.~Hern\'andez-Monteagudo\altaffilmark{17},
G.~C.~Hilton\altaffilmark{15},
M.~Hilton\altaffilmark{18,19},
A.~D.~Hincks\altaffilmark{10},
R.~Hlozek\altaffilmark{16},
K.~M.~Huffenberger\altaffilmark{20},
D.~H.~Hughes\altaffilmark{21},
J.~P.~Hughes\altaffilmark{22},
L.~Infante\altaffilmark{3},
K.~D.~Irwin\altaffilmark{15},
M.~Kaul\altaffilmark{5},
J.~Klein\altaffilmark{5},
A.~Kosowsky\altaffilmark{12},
J.~M.~Lau\altaffilmark{23,24,10},
M.~Limon\altaffilmark{25,5,10},
R.~H.~Lupton\altaffilmark{1},
K.~Martocci\altaffilmark{26,10},
P.~Mauskopf\altaffilmark{8},
F.~Menanteau\altaffilmark{22},
K.~Moodley\altaffilmark{18,19},
H.~Moseley\altaffilmark{13},
C.~B.~Netterfield\altaffilmark{27},
M.~D.~Niemack\altaffilmark{15,10},
L.~A.~Page\altaffilmark{10},
L.~Parker\altaffilmark{10},
H.~Quintana\altaffilmark{3},
B.~Reid\altaffilmark{28,10},
N.~Sehgal\altaffilmark{23},
B.~D.~Sherwin\altaffilmark{10},
J.~Sievers\altaffilmark{6},
D.~N.~Spergel\altaffilmark{1},
S.~T.~Staggs\altaffilmark{10},
D.~S.~Swetz\altaffilmark{5,15},
E.~R.~Switzer\altaffilmark{26,10},
R.~Thornton\altaffilmark{5,29},
H.~Trac\altaffilmark{30,1},
C.~Tucker\altaffilmark{8},
R.~Warne\altaffilmark{18},
G.~Wilson\altaffilmark{31},
E.~Wollack\altaffilmark{13},
Y.~Zhao\altaffilmark{10}
}
\altaffiltext{1}{Department of Astrophysical Sciences, Peyton Hall, 
Princeton University, Princeton, NJ USA 08544}
\altaffiltext{2}{Current address: Dept. of Physics and Astronomy, The Johns Hopkins University, 3400 N. Charles St., Baltimore, MD 21218-2686}
\altaffiltext{3}{Departamento de Astronom{\'{i}}a y Astrof{\'{i}}sica, 
Facultad de F{\'{i}}sica, Pontific\'{i}a Universidad Cat\'{o}lica de Chile,
Casilla 306, Santiago 22, Chile}
\altaffiltext{4}{Institute for the Physics and Mathematics of the Universe, 
The University of Tokyo, Kashiwa, Chiba 277-8568, Japan}
\altaffiltext{5}{Department of Physics and Astronomy, University of
Pennsylvania, 209 South 33rd Street, Philadelphia, PA, USA 19104}
\altaffiltext{6}{Canadian Institute for Theoretical Astrophysics, University of
Toronto, Toronto, ON, Canada M5S 3H8}
\altaffiltext{7}{Department of Physics and Astronomy, Haverford College,
Haverford, PA, USA 19041}
\altaffiltext{8}{School of Physics and Astronomy, Cardiff University, The Parade, Cardiff, Wales, UK CF24 3AA}
\altaffiltext{9}{Department of Physics and Astronomy, University of
British Columbia, Vancouver, BC, Canada V6T 1Z4}
\altaffiltext{10}{Joseph Henry Laboratories of Physics, Jadwin Hall,
Princeton University, Princeton, NJ, USA 08544}
\altaffiltext{11}{Department of Physics, University of Rome ``La Sapienza'', 
Piazzale Aldo Moro 5, I-00185 Rome, Italy}
\altaffiltext{12}{Department of Physics and Astronomy, University of Pittsburgh, 
Pittsburgh, PA, USA 15260}
\altaffiltext{13}{Code 553/665, NASA/Goddard Space Flight Center,
Greenbelt, MD, USA 20771}
\altaffiltext{14}{Berkeley Center for Cosmological Physics, LBL and
Department of Physics, University of California, Berkeley, CA, USA 94720}
\altaffiltext{15}{NIST Quantum Devices Group, 325
Broadway Mailcode 817.03, Boulder, CO, USA 80305}
\altaffiltext{16}{Department of Astrophysics, Oxford University, Oxford, 
UK OX1 3RH}
\altaffiltext{17}{Max Planck Institut f\"ur Astrophysik, Postfach 1317, 
D-85741 Garching bei M\"unchen, Germany}
\altaffiltext{18}{Astrophysics and Cosmology Research Unit, School of
Mathematical Sciences, University of KwaZulu-Natal, Durban, 4041,
South Africa}
\altaffiltext{19}{Centre for High Performance Computing, CSIR Campus, 15 Lower
Hope St. Rosebank, Cape Town, South Africa}
\altaffiltext{20}{Department of Physics, University of Miami, Coral Gables, 
FL, USA 33124}
\altaffiltext{21}{Instituto Nacional de Astrof\'isica, \'Optica y 
Electr\'onica (INAOE), Tonantzintla, Puebla, Mexico}
\altaffiltext{22}{Department of Physics and Astronomy, Rutgers, 
The State University of New Jersey, Piscataway, NJ USA 08854-8019}
\altaffiltext{23}{Kavli Institute for Particle Astrophysics and Cosmology, Stanford
University, Stanford, CA, USA 94305-4085}
\altaffiltext{24}{Department of Physics, Stanford University, Stanford, CA, 
USA 94305-4085}
\altaffiltext{25}{Columbia Astrophysics Laboratory, 550 W. 120th St. Mail Code 5247,
New York, NY USA 10027}
\altaffiltext{26}{Kavli Institute for Cosmological Physics, 
Laboratory for Astrophysics and Space Research, 5620 South Ellis Ave.,
Chicago, IL, USA 60637}
\altaffiltext{27}{Department of Physics, University of Toronto, 
60 St. George Street, Toronto, ON, Canada M5S 1A7}
\altaffiltext{28}{ICREA \& Institut de Ciencies del Cosmos (ICC), University of
Barcelona, Barcelona 08028, Spain}
\altaffiltext{29}{Department of Physics , West Chester University 
of Pennsylvania, West Chester, PA, USA 19383}
\altaffiltext{30}{Harvard-Smithsonian Center for Astrophysics, 
Harvard University, Cambridge, MA, USA 02138}
\altaffiltext{31}{Department of Astronomy, University of Massachusetts, 
Amherst, MA, USA 01003}
\altaffiltext{32}{Southern African Astronomical Observatory, Observatory Road,
Observatory 7925, South Africa}

\begin{abstract}
We report on extragalactic sources detected in a 455 square-degree map of the southern sky made with data at a frequency of \arone\ from the Atacama Cosmology Telescope 2008 observing season. We provide a catalog of \numberOfSources\ sources with flux densities spanning two orders of magnitude: from \minimumFlux\ to \maximumFlux. Comparison to other catalogs shows that 98\% of the ACT detections correspond to sources detected at lower radio frequencies. Three of the sources appear to be associated with the brightest cluster galaxies of low redshift X-ray selected galaxy clusters. Estimates of the radio to mm-wave spectral indices and differential counts of 
the sources further bolster the hypothesis that they are nearly all radio sources, and that their emission is not dominated by re-emission from warm dust. In a bright ($>50$~mJy)  \arone-selected sample with complete cross-identifications from the Australia Telescope 20\,GHz survey, we observe an average steepening of the spectra between 5, 20, and \arone\  with median spectral indices of $\alpha_{\rm 5-20} = -0.07 \pm 0.06$, $\alpha_{\rm 20-148} = -0.39 \pm0.04$, and $\alpha_{\rm 5-148} = -0.20 \pm 0.03$. When the measured spectral indices are taken into account, the \arone\ differential source counts are consistent with previous measurements at 30 GHz in the context of a source count model dominated by radio sources. Extrapolating with an appropriately rescaled model for the radio source counts, the Poisson contribution to the spatial power spectrum from synchrotron-dominated sources with flux density less than 20~mJy is $C^{\rm Sync} = (2.8 \pm 0.3) \times 10^{-6} \micro\kelvin^2$.
\end{abstract}

\keywords{surveys --- radio continuum: galaxies --- galaxies: active --- cosmic background radiation}


\section{INTRODUCTION}

\setcounter{footnote}{0}

Large ($> 100$\,square-degrees) millimeter-wave surveys are beginning to probe arcminute angular scales,  corresponding to spatial frequencies $\ell > 3000$. At these small angular scales, the fluctuations in the extragalactic sky temperature become dominated by emission from galaxies and the thermal Sunyaev-ZelÕdovich (SZ) effect from galaxy clusters \citep{sunyaev/zeldovich:1970}, rather than primordial fluctuations in the Cosmic Microwave Background (CMB). 

 The predominant extragalactic point sources of emission at  \arone\ (2.0\,mm) are active galactic nuclei (AGN) and dusty, star-bursting galaxies. AGN detected at \arone\ are characterized by a synchrotron-dominated spectrum extending to lower radio frequencies. On the other hand, dusty, star-bursting galaxies at \arone\ display a grey body spectrum increasing with frequency into the submillimeter.  The source of \arone\ flux from AGN is synchrotron emission concentrated near the central accreting super massive black hole while for dusty star-bursting galaxies the millimeter flux is sourced by re-emission from dust that is heated primarily by prodigious star formation. While the millimeter emission from the majority of dusty star-bursting galaxies is below the nominal flux  density limits of current large-scale surveys, a recent 150 and 220\,GHz study by  \cite{vieira/etal:2010} using the South Pole Telescope identifies a sub-population of these sources with anomalously high fluxes which likely belong to a rare, lensed population of high-redshift dusty galaxies \citep{negrello/etal:2007,lima/etal:prep}. It follows that current and future wide-area millimeter surveys will identify important sub-populations of core-exposed, radio-loud AGN and lensed dusty galaxies. With the members of these sub-populations identified, detailed follow-up studies will help us better understand these high-energy states of galaxy formation as well as their important role in providing energy feedback to their environments.

Because source emission is a significant contributor to the overall sky brightness at small scales, the characterization of extragalactic sources is essential for interpreting the primary CMB anisotropies and the SZ signal from galaxy clusters. Measurements of the primary CMB power spectrum at high spatial frequencies ($\ell \geq 2000$) will constrain the form of the inflationary potential (e.g., the spectral index $n_s$ of primordial fluctuations). Such measurements will require information about the spectral and spatial distribution of millimeter sources in order to separate foregrounds from the primordial signal. At still smaller scales ($\ell > 3000$), studies of the CMB spectrum attempt to constrain fluctuations in the matter density field  from the contribution of the SZ to the power spectrum \citep{lueker/etal:2010, fowler/etal:2010}. For these studies an in-depth understanding of the point source populations is even more critical for separating the power spectrum of the SZ from that of sources. Furthermore, an understanding of the energy feedback from AGN and star-formation to the cluster environment will be important for constraining the form of the SZ spectrum \citep{battaglia/etal:prep,shaw/etal:prep,trac/bode/ostriker:prep}. Finally, SZ surveys attempting to measure $\Omega_{\rm M}$, $\sigma_8$, and dark energy through the evolution of the cluster mass function will likewise need to consider the spectral behavior and cluster occupation numbers for sources in order to avoid systematically biasing mass estimates based on SZ flux density \citep{lin/etal:2009,sehgal/etal:2010,vanderlinde/etal:prep}.

The Atacama Cosmology Telescope (ACT) is a millimeter-wave observatory which will ultimately survey thousands of square degrees of sky at arcminute resolution with milli-Jansky sensitivity to sources. ACT is located at 5200\,m in the Atacama Desert in the Andes of northern Chile.\footnote{The ACT Site is at 22.9586$\degree$ south latitude, 67.7875$\degree$ west longitude.} This high desert site in the tropics was chosen for its excellent atmospheric transparency and its access to northern and southern celestial latitudes. ACT observes simultaneously in bands centered at \arone\ (2.0\,mm), \artwo\ (1.4 mm), and \arthree\ (1.1 mm), each band having a dedicated 1024-element array of bolometric transition edge sensors. As of mid-2010, ACT has completed three seasons of observations: 2007, 2008, and 2009, and the 2010 season is underway. In each season, ACT has conducted two surveys: a $9\degree$-wide survey centered at declination $-53.5\degree$ and a $5\degree$-wide stripe centered on the celestial equator. 

In this paper we report on extragalactic sources in the ACT 2008 \arone\ dataset. This is the first report devoted to ACT source science and complements the \arone\ power spectrum study in \citet{fowler/etal:2010}. In what follows we give an overview of the observations and data reduction (Section \ref{sec:data}), describe the source catalog (Section \ref{sec:catalog} and the Appendix), and discuss implications of the study including constraints on source models (Section \ref{sec:discussion}).

\begin{figure*}[ht]
\begin{center}
  \resizebox{\textwidth}{!}{
    \plotone{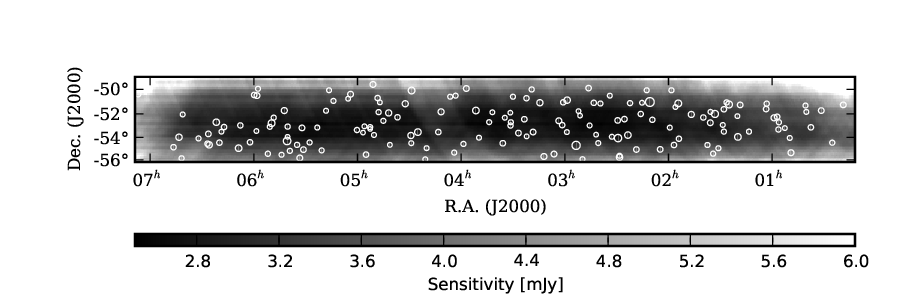}
  }
\caption{Sensitivity map with detections. The subset of the ACT 2008 \arone\ dataset  considered for this study lies between right ascension $00^{\mathrm h}12^{\rm m}$ and $07^{\mathrm h}08^{\rm m}$ and declination $-56\degree11\arcmin$ and $-49\degree00\arcmin$ (\area). The gray-scale encodes the rms of the map in mJy. The deepest data correspond to an exposure time of 23.5 minutes per square-arcminute and a 1$\sigma$ sensitivity of 2.5 mJy. White circles mark the locations of ACT sources. The diameter of each circle is proportional to the log of the associated source flux density. Towards the edge of the map, the noise properties display local variation. For this reason, detections with flux density values below 50 mJy have been discarded in regions where the rms exceeds 4.6 mJy, corresponding to less than 7 minutes of integration. This, together with an exclusion of all detections below \significance\  $\sigma$, accounts for the relative dearth of detections in areas of shallow coverage (See Section \ref{subsec:sourceDetection}.). }
\label{fig:coverage}
\end{center}
\end{figure*}

\section{Observations and Data Reduction}
\label{sec:data}
The data used for this analysis were collected by ACT at \arone\ during its second observing season in 2008. This section gives an overview of the survey observations and the reduction of the raw data to a map as well as a detailed description of the source extraction. For a more thorough introduction to the ACT facility, observations, and data reduction pipeline, we refer the reader to \citet{fowler/etal:2010}, \citet{swetz/etal:prep} and references therein.\footnote{ACT Collaboration papers are archived at http://www.physics.princeton.edu/act/.} 

\subsection{Observations}
\label{subsec:observations}
The 2008 southern observations were carried out over a survey area $9\degree$ wide, centered on declination  $-53.5\degree$, and extending from right ascension $19^{\mathrm h}$ to $24^{\mathrm h}$ and $00^{\mathrm h}$ to $07^{\mathrm h}30^{\mathrm m}$. The subset of these data used in the present analysis lies between right ascensions $00^{\mathrm h}12^{\mathrm m}$ and $07^{\mathrm h}08^{\mathrm m}$ and declinations $-56\degree11\arcmin$ and $-49\degree$00$\arcmin$ (\area). The area was chosen to encompass the data used for power spectrum work in \citet{fowler/etal:2010} and represents a large fraction of the most deeply covered regions from the 2008 \arone\ dataset. Figure \ref{fig:coverage} shows the area of the sky used and associated point source sensitivies. Typical white noise levels in the map are 30-50 $\micro\kelvin$-arcminute, tending to higher values towards the map boundaries. As described in Section  \ref{subsec:sourceDetection}, this white noise level, when match filtered with the ACT beam, results in typical sensitivities to point source flux densities from 2.5 to 5 mJy.

The 2008 ACT observing season extended from mid-August to the final week of December. Observations took place during nighttime hours: from roughly 20:00 to 06:00 local time. Of the total observing time, approximately 85\% was devoted to the southern region. ACT observed by scanning at a constant elevation of 50.5$\degree$ while the survey region drifted through the scan with the rotation of the Earth. During the first half of a night, ACT scanned at azimuth 150$\degree$, targeting a rising section of the survey area. During the second half of the night, ACT scanned the same section setting on the other side of the south celestial pole at azimuth 210$\degree$. The rising scans cross the survey region from southwest to northeast and back (in equatorial coordinates), while the setting scans cross the survey region from southeast to northwest. Together, the rising and setting scans cross-link each point on the sky with all adjacent points.  The resulting cross-linked temperature data in principle contain all information necessary to recover an unbiased, low-noise map of the millimeter sky. In addition to survey observations, ACT also executed regular observations of Uranus and Saturn during 2008 to provide gain calibration, beam profiles, and pointing.

With the telescope scan strategy described above, any given location in the survey area would be observed over a period of approximately two months during a season. Therefore, source flux densities reported here are the average flux density over a two month period. This is an important point as the vast majority of sources presented in this paper are AGNs which are known to be variable \cite[e.g.,][]{kesteven/etal:1977,valtaoja/etal:1992}.

\begin{figure*}[ht]
\begin{center}
  \resizebox{\textwidth}{!}{
    \plotone{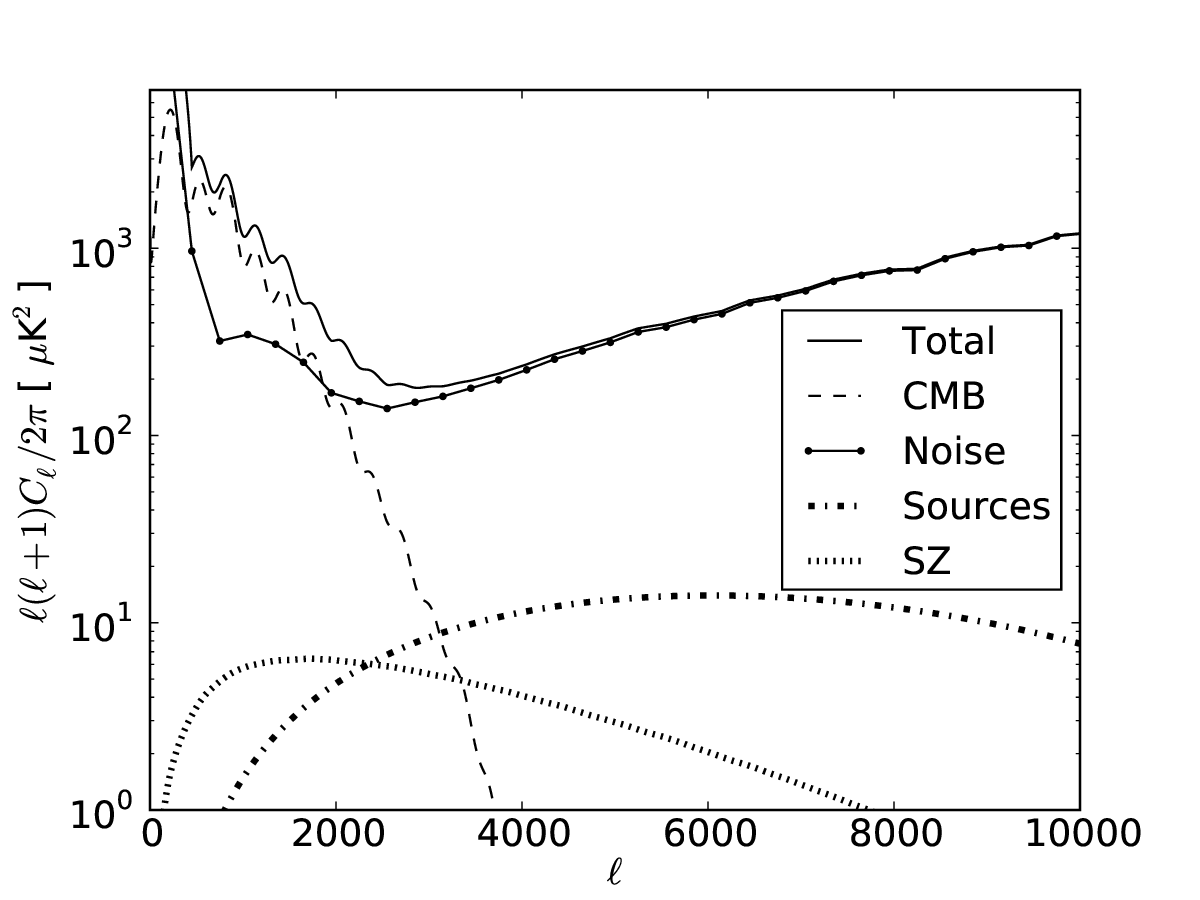}
        \plotone{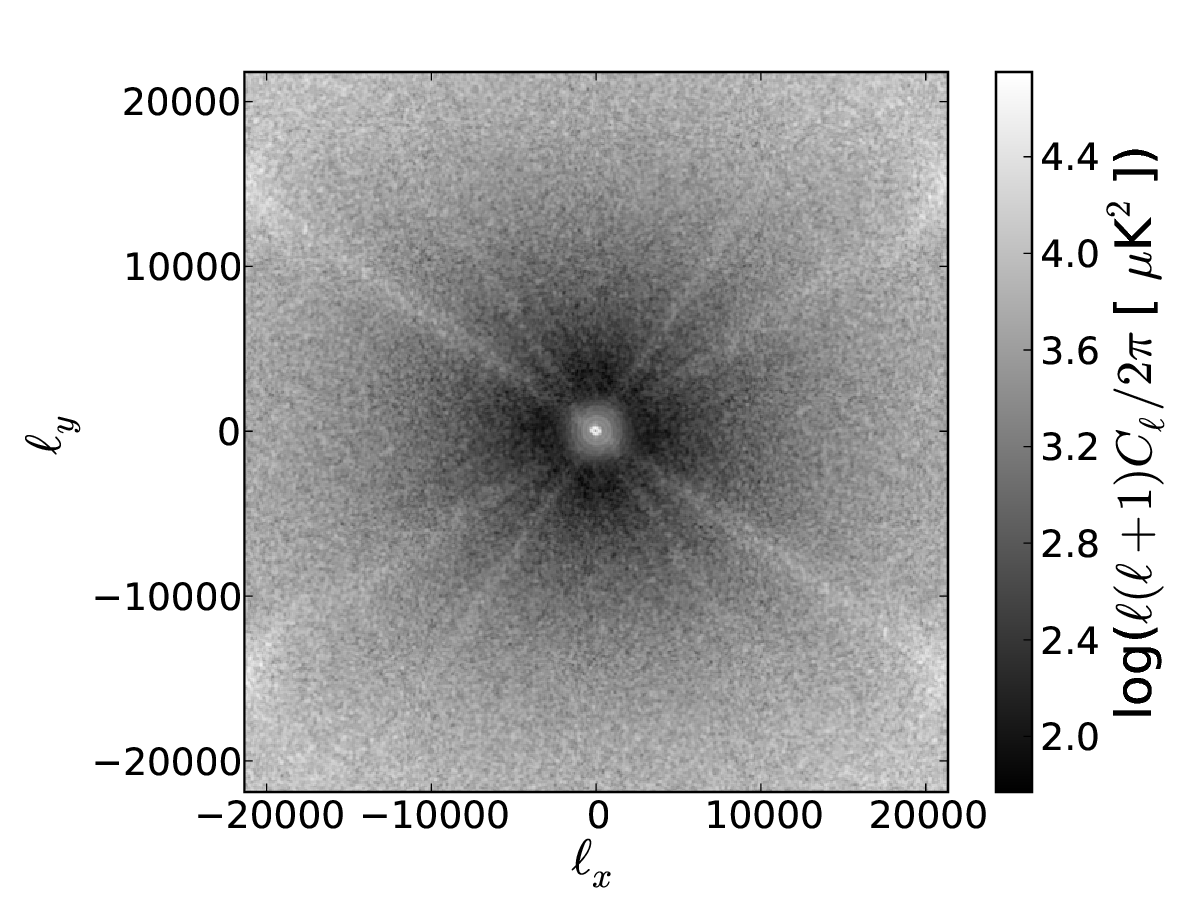}
  }
  
\caption{Model ACT auto-power spectra used for the matched filter. \emph{Left.} One-dimensional spectra decomposed by component. The CMB spectrum is taken from WMAP5 \citep{nolta/etal:2009}. The undetected sources and the Sunyaev-Zel'dovich effect from galaxy clusters are 1D templates fit to the high-$\ell$ \arone\ spectrum in \citet{fowler/etal:2010}. The 1D noise spectrum is obtained by radially-binning the average of 2D spectra from ACT \arone\ jackknife maps. The spectra are dominated at low-$\ell$ by the CMB and atmospheric noise and at high-$\ell$ by white detector and photon shot noise. Convolution of  all celestial components by the ACT beam results in attenuation of the corresponding spectra at high-$\ell$.  \emph{Right.} The model two-dimensional auto power spectrum. The spectrum includes noise, CMB, undetected sources and the Sunyaev-Zel'dovich effect from galaxy clusters. The latter three components were obtained from the 1D models. The ACT noise is isotropic except for extra noise in scan parallel (from residual 1/f) and scan perpendicular (from detector row correlations) directions. For each rising and setting scan direction, two orthogonal bands of excess power are centered on the origin rotated $60^\degree$ with respect to one another. To minimize contamination, it is important to deweight data in these diagonal modes through the two-dimensional matched filter.}
\label{fig:power}
\end{center}
\end{figure*}

\subsection{Reduction to Maps}
\label{subsec:reduction}
The raw \arone\ data consist of 1024 time-ordered data streams, one per element of the detector array. Approximately 25\% of the data are rejected on the basis of telescope operation and weather. After further cuts based on individual detector performance, the data from 680 \arone\ detectors over 850 hours ($\sim 3200$ GB) are retained from the 2008 southern survey. 

Pointing reconstruction is accomplished in two steps. First, the relative detector pointings are established with 1.2$''$ certainty through observations of Saturn. Second, absolute detector array pointings for our two southern survey configurations (rising and setting: 150$\degree$ and 210$\degree$  azimuth, 50.5$\degree$ elevation) are established with $3.5\arcsec$ precision through an iterative process in which the absolute pointing is adjusted based on offsets of ACT-observed radio source locations with respect to source locations taken from the Australia Telescope 20\,GHz (AT20G) survey \citep{murphy/etal:2010}.

Nightly calibrations of the detectors' responsivity (power-to-current conversion) are based on load curves taken at the start of each night. Stability of this calibration through the night is monitored using small steps in the detector bias voltages and established at the few percent level.   Relative detector flux density calibrations are based on normalizing the detector responses to the beam-filling atmospheric signal. The resulting relative calibration is shown to be constant through the season at the few percent level. The final brightness temperature calibration is based on ACT observations of Uranus throughout the season and the WMAP7 Uranus temperature \citep{weiland/etal:inprep} extrapolated to the ACT \arone\ band. The calibration is more fully described in \citet{fowler/etal:2010}. The overall calibration is certain to 6\% rms, a number dominated by systematic uncertainties in extrapolating the temperature of Uranus to \arone\ from WMAP frequencies. 

The final step in the data reduction is map-making. An iterative preconditioned conjugate gradient solver is used to recover the maximum likelihood (ML) maps. The algorithm solves simultaneously for the millimeter sky as well as correlated noise (e.g., a common mode from atmospheric emission).  The map projection used is cylindrical equal area with a standard latitude of $-53.5\degree$ and $0.5\arcmin$ square pixels. For more details on the mapping and other reduction steps, refer to \citet{fowler/etal:2010}.

\subsection{Data Modeling}
\label{subsec:dataModeling}
The next step toward a source catalog is the construction of a filter which optimizes the signal-to-noise ratio (S/N) of sources in the \arone\ map.\footnote{See Section \ref{subsec:sourceDetection} for the definition of S/N used in this work.} In order to construct such a filter it is necessary to obtain best estimates of the power spectra of the different components, signal and noise, which contribute to the ACT data. We model the temperature $T$ at position $\vec{x}$ as the sum over sources plus other components in the map:
\begin{equation}
T(\vec{x}) = \sum_i T_i b(\vec{x}-\vec{x_i}) + T_{\rm other}(\vec{x})
\end{equation}
where $T_i$ is the peak amplitude of the $i^{\rm th}$ source, $b$ is the ACT \arone\ beam function normalized to unit amplitude and taken to be isotropic \citep{hincks/etal:prep}, and $T_{\rm other}$ includes contributions from the primary CMB, undetected point sources, SZ from clusters and noise from the detectors and atmosphere.

\begin{figure*}[ht]
\begin{center}
  \resizebox{\textwidth}{!}{
    \plotone{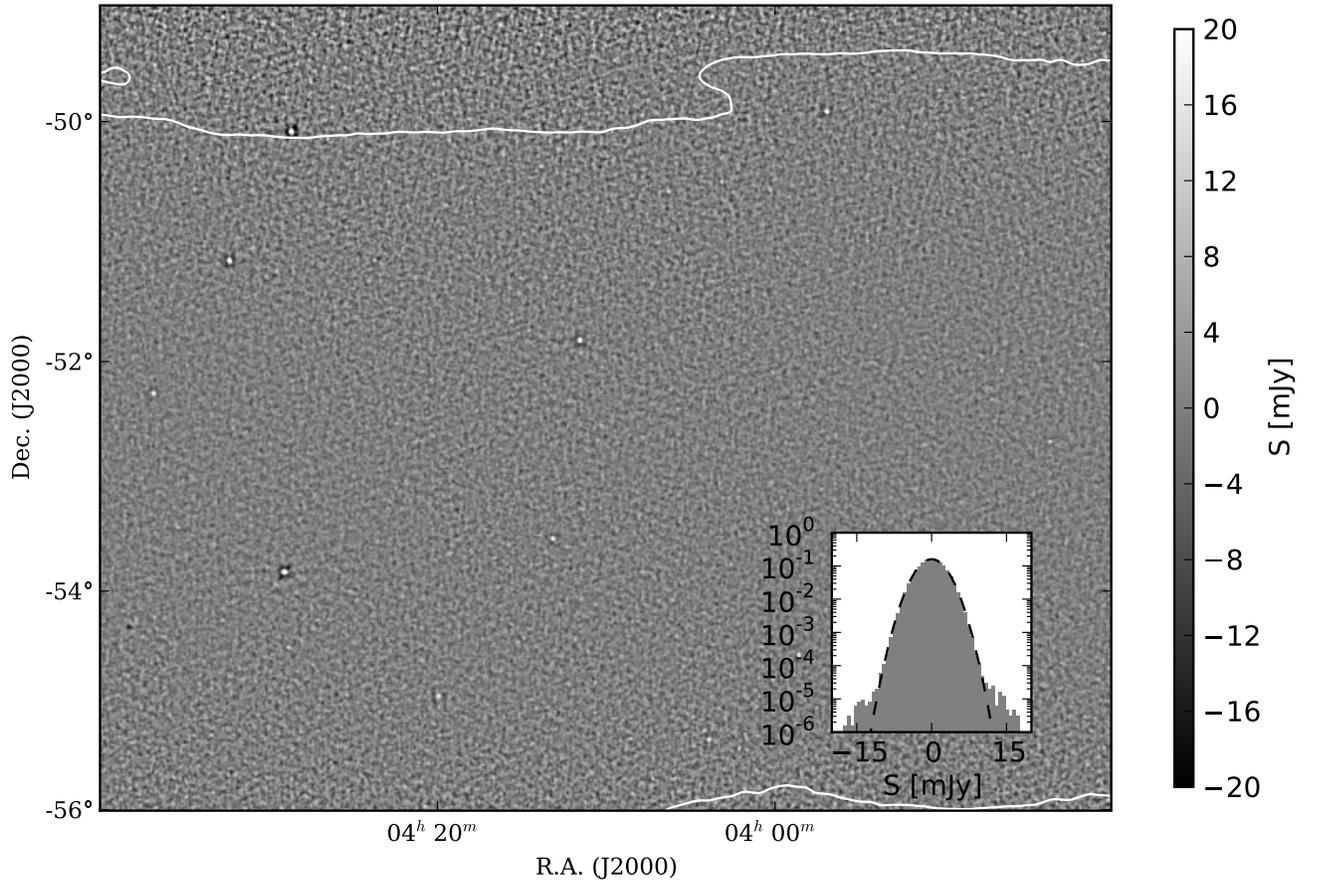}
  }
\caption{\arone\ map. The submap shown above is a sample 64 square-degrees of the survey field. The data have been matched filtered such that the gray-scale is in units of flux density (mJy). The inset shows the flux density distribution across the data weighted by the max-normalized square-root of the number of data per pixel  $\sqrt{N_{\rm obs}(\vec{x})}/\sqrt{N_{\rm obs, max}}$. Thus the distribution represents the data in the deepest part of the map (although it uses weighted data from all regions). The data distribution is shown as a grey histogram on which is plotted a dashed Gaussian distribution with standard deviation 2.5 mJy. The positive non-Gaussian tail may, in part, be attributed to sources and the negative tail to ringing from the filter about sources as well as SZ (e.g., ACT-CL J0438-5419 from \citet{menanteau/etal:2010}). Several sources are apparent as white points surrounded by darker rings from the filter. The white contour marks the transition at the edge of the map where the rms exceeds 4.6 mJy, corresponding to less than 7 minutes of integration per arcminute. In this region we have excluded detections below 50 mJy due to contamination from local noise.  The source above the contour at $\delta \approx -50\degree$ has a flux density of 150 mJy and is included in the catalog. }
\label{fig:map}
\end{center}
\end{figure*}

Figure \ref{fig:power} shows graphs of the power spectra of the contributors to $T_{\rm other}$. For the primary CMB component we use the WMAP5 best fit model from \citet{nolta/etal:2009}. The SZ and undetected source components are taken from models fit to the ACT \arone\ power spectrum in \citet{fowler/etal:2010}. The spatial power spectrum from \citet{fowler/etal:2010} was computed after masking sources from a preliminary version of the catalog presented in this paper. As such the derived contribution to $T_{\rm other}$ from undetected sources should be a suitable estimate for the matched filter in this work. Specifically, the source model is a Poisson spectrum normalized to $\ell(\ell+1)C_p/2\pi = 11.2$~$\micro {\rm K}^2$ at $\ell = 3000$, and the SZ model is a combined thermal and kinetic SZ model ($\sigma_8$ = 0.8) from \cite{sehgal/etal:2010} normalized by a best-fit factor of 0.63. These celestial models (CMB, sources, and SZ) are all convolved with the ACT \arone\ beam function.  The celestial models are natively one-dimensional functions of multipole $\ell$ (as shown in the left graph in Figure \ref{fig:power}) with corresponding azimuthally symmetric two-dimensional representations (as shown in the right graph in Figure \ref{fig:power}). The noise term, on the other hand, is natively a two-dimensional power spectrum estimated by the average of power spectra from difference maps. The difference maps were constructed by subtracting a map made from one half of the data (with a random selection of observing days contributing) from a map made from the other half (using the remaining observing days). The two-dimensional noise power spectrum is binned radially to obtain the one-dimensional representation shown by the line of connected dots in the left graph of Figure \ref{fig:power}. 

From Figure  \ref{fig:power} it is clear that the primary contributions to $T_{\rm other}$ come from the CMB and atmosphere at large scales ($\ell < 2500$) while the white detector and photon shot noise dominates the power at small scales. In the spectral trough around $\ell = 2500-3500$ the detectable sources will have their greatest signal-to-noise ratio. 

An important feature of the two-dimensional power spectrum in the right graph of Figure \ref{fig:power} is the anisotropic nature of the noise term which corresponds to striping in the map. The stripes  are a result of large scale drifts in atmospheric emission along the scan directions as well as from correlations among rows of ACT detectors perpendicular to the scan directions. Accounting for this anisotropy when filtering the map is important for extracting an uncontaminated sample of sources. To properly down-weight these noisy diagonal modes, we adopt a two dimensional noise covariance in the matched filtering technique described in Section \ref{subsec:sourceDetection}.

The final significant noise term that is not captured in difference maps is consistent with scan synchronous noise. The noise is likely due to instabilities induced by acceleration at scan turn-arounds. The noise manifests itself as horizontal striping in the maps and as excess power isolated to a vertical strip $-100 < \ell_x < 100$ in Fourier space. For simplicity, we set the noise power to infinity in this section of Fourier space such that modes contaminated by this noise will be completely down-weighted by the filter as described in the next section.

\subsection{Detection and Catalog Construction}
\label{subsec:sourceDetection}

To optimize the S/N of the sources with respect to the  background, we use a matched filter. This approach has been proposed and used in previous work to find both sources \citep{tegmark/deOliveira-Costa:1998,wright/etal:2009,vieira/etal:2010} and the Sunyaev-Zel'dovich effect from clusters \citep{melin/bartlett/delabrouille:2006,staniszewski/etal:2009}. For completeness of presentation, here we rederive the form of the filter and its basic properties. Without loss of generality, we consider a field centered on a source such that
\begin{equation}
T(\vec{x}) =  T_0 b(\vec{x}) + T_{\rm other}(\vec{x}).
\label{equ:model}
\end{equation}
We apply a filter $\Phi(\vec{k})$ in Fourier space such that the temperature at the center of filtered field is
\begin{equation}
T_{\rm filt}(0) = T_0 \int \Phi(\vec{k}) \tilde{b}(\vec{k}) {\rm d}\vec{k}  + \int \Phi(\vec{k})\widetilde{T}_{\rm other}(\vec{k})  {\rm d}\vec{k}
\end{equation}
where $\tilde{b}$ and $\widetilde{T}_{\rm other}$ are the Fourier transform of the beam and noise temperature field, respectively.  The  S/N of the central source is
\begin{equation}
{\rm S/N} = \frac{\mid T_0 \int \Phi(\vec{k}) \tilde{b}(\vec{k}) {\rm d}\vec{k} \mid}{ \mid \int \Phi(\vec{k})\widetilde{T}_{\rm other}(\vec{k})  {\rm d}\vec{k} \mid}.
\end{equation}
A filter that maximizes this S/N is
\begin{equation}
\Phi(\vec{k}) = \frac{  \tilde{b}^*(\vec{k}) \mid \widetilde{T}_{\rm other}(\vec{k}) \mid^{-2}} { \int   \tilde{b}^*(\vec{k'}) \mid \widetilde{T}_{\rm other}(\vec{k'}) \mid^{-2} \tilde{b}(\vec{k'}) {\rm d}\vec{k'}}
\label{equ:filter}
\end{equation}
where the normalization has been chosen to produce an unbiased estimate of the amplitude of the source $T_0$ at $\vec{x} = 0$. The noise variance of the filtered data is
\begin{eqnarray}
\sigma^2 & = & \int \mid \Phi(\vec{k}) \widetilde{T}_{\rm other}(\vec{k}) \mid^2 {\rm d}\vec{k} \nonumber \\
 & = & \left[ \int  \tilde{b}^*(\vec{k}) \mid \widetilde{T}_{\rm other}(\vec{k}) \mid^{-2} \tilde{b}(\vec{k}) {\rm d}\vec{k}\right]^{-1} .
 \label{eqn:noiseVariance}
\end{eqnarray}
In practice, the noise variance of the filtered map is obtained from the filtered map itself after masking the brightest six sources (S/N $> 50$). These sources increase the rms of the maps by 10\%. The remaining sources contribute approximately 1\% to the rms. This slightly more conservative estimate of the variance agrees with the estimate from Equation \ref{eqn:noiseVariance} in which $T_{\rm other}$ has been constructed as described in Section \ref{subsec:dataModeling}.  In what follows, the S/N at a location in the filtered map is defined as the temperature at that location divided by the square-root of this variance. 

Before applying the global matched filter from Equation \ref{equ:filter}, we multiply the map, pixel-wise, by the square-root of the number of observations per pixel normalized by the observations per pixel in the deepest part of the map, $\sqrt{N_{\rm obs}(\vec{x})/N_{\rm obs, max}}$. This is equivalent to weighting the data by the inverse of the estimated white noise rms shown in Figure \ref{fig:coverage} and accounts for local variation in the white noise amplitude. Furthermore, the map is tapered to zero in a 10$\arcmin$ boundary region around the edge of the map to mitigate the artifacts arising from data aperiodicity when filtering. This windowed data is excluded from the final analysis, reducing the usable sky area from the total 455 square-degrees to 443 square-degrees. Next, because the ringing of the filter around the brightest sources can cause false detections, we identify and mask the six most significant ( $> 50 \sigma$ ) sources before applying the matched filter to the map. These brightest sources are treated and included in the catalog in the same manner as the fainter sources with the exception they are recovered through an initial run of the pipeline with the S/N lower limit increased.  An important final caveat: in constructing the noise term $\widetilde{T}_{\rm other}$, the component models need to be tapered and weighted in the same fashion as the data for the matched filter formalism to hold. This is particularly true for components with red spatial spectra, such as the atmosphere and CMB, because aliasing due to a particular windowing scheme can significantly alter the spectrum. Figure \ref{fig:map} shows a sample 64 square-degrees of the filtered map. For this reason the catalog includes only detections with S/N \significance\ and above. See Section \ref{subsec:purityAndCompleteness} for a discussion of purity determination.

Localized, non-white noise in the map requires that we take further measures beyond the global matched filter solution outlined above. First, local large-scale atmospheric noise requires us to add a low-$\ell$ taper to the term $\mid \widetilde{T}_{\rm other}(\vec{k}) \mid^{-2}$ (Equation \ref{equ:filter}) which rises from zero at $\ell = 0$ to one at $\ell = 1200$ as $\sin^5(\pi \ell/2400)$. This filter removes the local atmospheric noise while down-weighting only a small fraction ($\sim$1\%) of the data containing source power in Fourier space. Second, in areas of the map which are particularly shallow, uneven coverage leads to excess striping. In these parts of the map the noise model described in Section \ref{subsec:dataModeling} is invalid and non-white noise remains even after the filter is applied. For this reason, we exclude sources from the catalog which are detected with flux densities below 50 mJy in areas of the map with exposure times less than 7 minutes per square-arcminute. This exposure time is approximately one-third the exposure time in the deepest areas of the map and corresponds to 4.6 mJy rms. The cut level of 50 mJy at 7 minutes per square-arcminute was chosen to broadly eliminate contamination observed in simulations. Future studies will make use  of local noise estimation to avoid such an exclusion. The sample submap in Figure \ref{fig:map} demarcates the region at the edge of the map in which we exclude 50 mJy detections with a white contour.  With this exclusion, the area used for sources with flux density below 50 mJy is 366 square-degrees.  

The final step in the catalog generation is to derive the flux densities associated with the detections. Given the form of the filter in Equation \ref{equ:filter}, the source-centered value of the filtered map, multiplied by the solid angle of the beam profile, is the source flux density. It is this value, rescaled by the inverse of the square-root of the number of observations per pixel normalized by the number of observations per pixel in the deepest part of the map ($\sqrt{N_{\rm max}/N_{\rm obs, max}(\vec{x})}$), that we record as the raw flux density estimate for a detection. It follows that an error in source location results in an error in flux estimation. Such an error arises due to finite pixel size: a detection rarely falls in the center of a pixel.  Source location error due to finite pixel size causes a systematic negative bias and increased scatter. The 0.5$\arcmin$ ACT map pixels cause a 10\% negative bias with comparable scatter in the flux density estimate. To remove this bias and scatter, we  zero-pad the filtered data in Fourier space (e.g., \citet{press/teukolsky/vetterling:NRC:2e}) such that the pixel spacing in map space is decreased by a factor of sixteen.  This Fourier interpolation of the filtered data onto 0.03125$\arcmin$ pixels is a convenient way to better locate the peak of source emission and therefore mitigate systematic errors in position and flux density estimation to less than 1\%.  An important caveat to this technique is that the pixel window function must be taken into account when reducing the pixel size. As discussed further in Section \ref{subsec:SimulationsAndFluxRecovery}, simulations show that flux densities thus derived are unbiased at the sub-percent level.

\begin{figure}[htbp]
\begin{center}
    \plotone{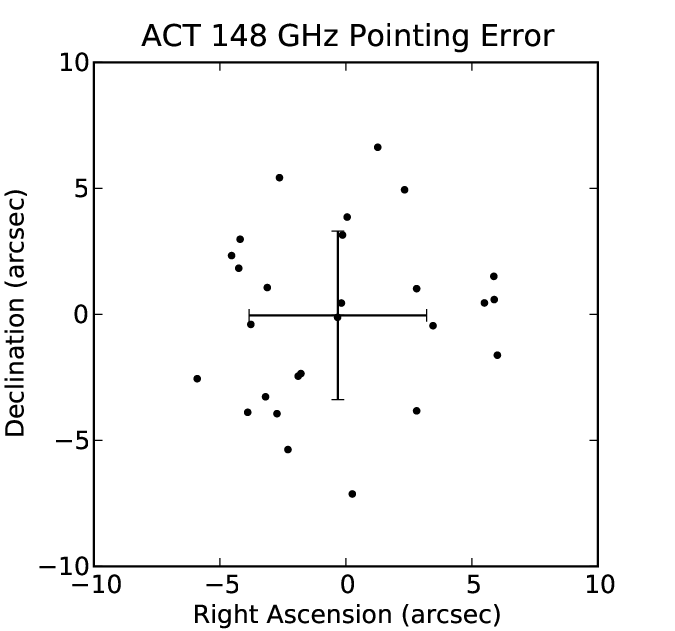}
\caption{Astrometric Accuracy. The small filled circles are the positional offsets of ACT sources with S/N $>$ 20 from
counterparts in AT20G. The error bars show the rms in right ascension (3.5$''$) and declination (3.3$''$) and are centered on the mean of the distribution:  $-0.3\arcsec\pm0.7\arcsec$ in right ascension and  $0.0\arcsec\pm0.6\arcsec$ in declination.}
\label{fig:pointing}
\end{center}
\end{figure}

\section{Catalog}
\label{sec:catalog}
The catalog of \numberOfSources\ ACT extragalactic sources is given in Table \ref{tab:sources} of the Appendix. The catalog provides the IAU name, celestial coordinates (J2000) as well as S/N and \arone\  flux density estimation of each ACT-detected source. Raw flux densities are estimated directly from the map as described in Section \ref{subsec:sourceDetection}. Deboosted flux densities (Section \ref{subsec:deboosting}) are also given.  The seventh column gives the exposure time in minutes per square-arcminute at the location of the source. Finally, we provide the ID of AT20G  sources collocated within 30$\arcsec$ of an ACT source. The following subsections provide the necessary information for interpreting the catalog data.

\subsection{Astrometric Accuracy}

The positions of the twenty-seven ACT sources with S/N $>$ 20 were compared to positions of associated sources in the AT20G catalog.  The AT20G pointing is checked against VLBI measurements of International Celestial Reference Frame calibrators and is shown to be accurate to better than one arcsecond \citep{murphy/etal:2010}. Figure \ref{fig:pointing} shows the offsets of the twenty-seven ACT source locations with respect to associated AT20G locations. The mean of the offsets is $-0.3\arcsec \pm 0.7\arcsec$ in right ascension and  $0.0\arcsec \pm 0.6\arcsec$ in declination. The rms of offsets is 3.5$\arcsec$ in right ascension and $3.3\arcsec$ in declination.  One ACT source with S/N $>$ 20 is cross-identified with an AT20G source $18\arcsec$ away. This source is ACT-S J042906-534943 and is flagged as extended in the AT20G catalog. Furthermore, there are multiple SUMSS sources located within the search radius suggesting a nontrivial geometry. Therefore, this source represents an outlier in this analysis and is not used. For twenty-four cross-identified sources with S/N $<$ 8, the ACT location rms with respect to AT20G position was inflated by the effect of the noise: $8.8\arcsec$ in right ascension and $6.2\arcsec$ in declination. 

We have identified a systematic shift in pointing throughout the season likely associated with movement of the telescope. A significant fraction of the scatter in Figure \ref{fig:pointing} is attributable to this systematic effect. Because of this effect,  the distribution of pointing errors in Figure \ref{fig:pointing} does not appear Gaussian. We plan to eliminate this effect in a future study.

\begin{figure}[htbp]
\begin{center}
    \plotone{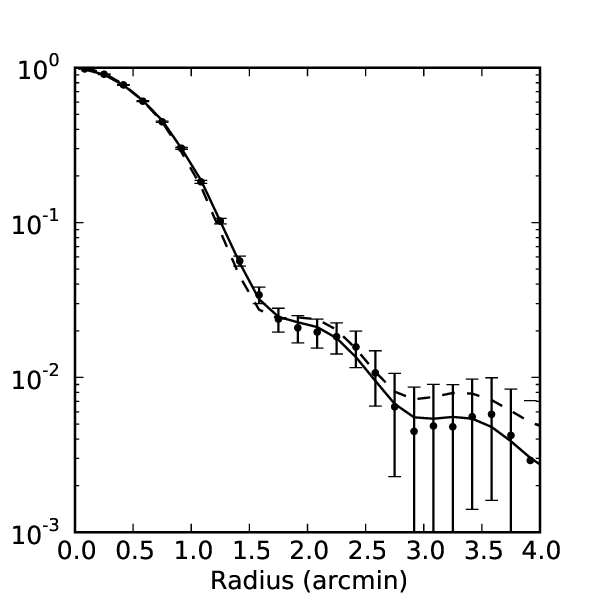}
\caption{Source Profile Fitting. The radially binned and normalized brightness profile observed for ACT-S J021046$-$510100 is represented by the points with error bars. The solid line running through the points is the measured ACT beam from \cite{hincks/etal:prep} convolved with a Gaussian with $\sigma=3''$ plus an additive background term. The Gaussian convolution is intended to represent spreading of the beam by random pointing error. The value of $\sigma=3''$ describes the best-fit to a Gaussian-convolved beam plus background term. Note that $3''$ is consistent with pointing uncertainty from Figure \ref{fig:pointing}. The dashed line shows the model with background and beam amplitude fit to the data, but no Gaussian convolution. Without the Gaussian convolution, the background term increases to fit the data at small angles and forces the model high at large angles. The data clearly prefer the Gaussian-convolved model. Similar fits to the twenty brightest sources in the ACT data suggest bias in the reported flux density level due to misestimation of the source profile is at the sub-percent level. The error bars shown on the radially binned points are correlated.}
\label{fig:beam}
\end{center}
\end{figure}

\subsection{Flux Density Recovery}
\label{subsec:fluxRecovery}
As discussed in Section \ref{subsec:reduction}, the overall calibration has an uncertainty of 6\%, dominated by systematic errors in the temperature of Uranus. In addition to this uncertainty, errors in the flux estimation may arise due to errors in the assumed source profile, to flux boosting of lower significance candidates, and to a failure of the map-maker to converge. In this section we describe tests of these potential sources of flux density error as well as an end-to-end check of the match-filter recovery and deboosting through simulations.

\subsubsection{Beam Profile}
\label{subsec:beam}
Flux density recovery is a function of the form of the source profile $b$ assumed in the filter (See Equations \ref{equ:model} and \ref{equ:filter}.).  For the filter's source profile $b$ we adopt the ACT \arone\ beam from \citet{hincks/etal:prep}. Deviation of the actual source profile from the ACT \arone\ beam will result in a biased estimate of the flux density as determined from the filtered map. 

\begin{figure}[ht]
\begin{center}
\plotone{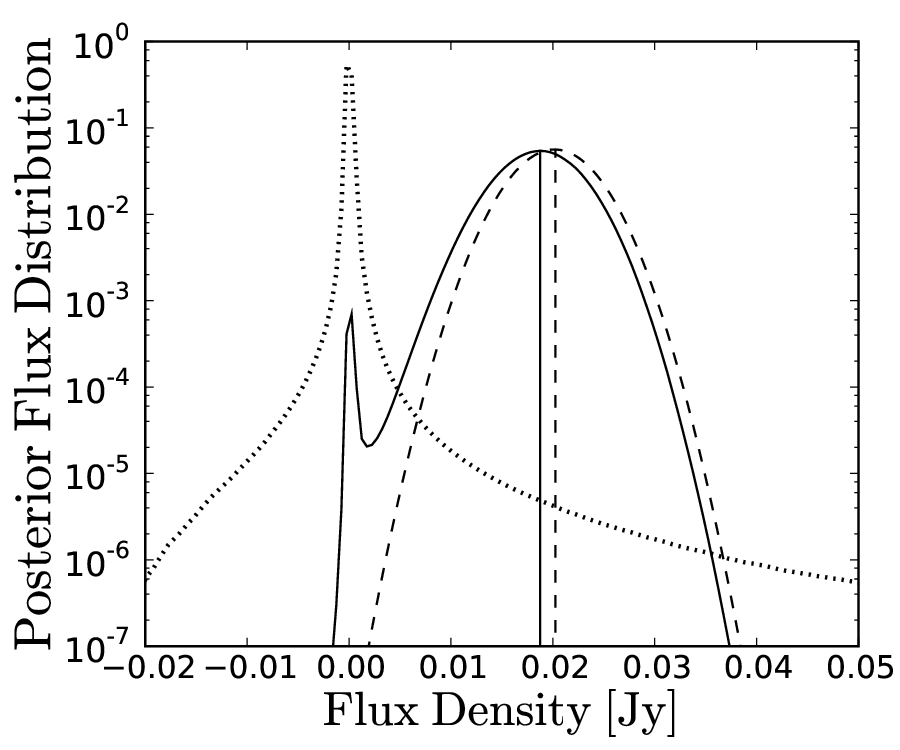}
\caption{ Flux density deboosting. The Gaussian flux density distribution ${P(S_m \mid S_0)}$ (dashed) derived directly from the maps is multiplied by the prior distribution ${P(S_0)}$ (dotted) of the intrinsic source flux to obtain the deboosted posterior flux distribution (solid). Since the mean of the maps is set to  zero through filtering, $P(S_0)$ can assume negative values. The peak in the distribution shifts to lower density flux (a $<$10\% effect for 20\,mJy sources). For the S/N=5.6 source shown here, there is little volume in the posterior ${P(S_m \mid S_0)}$ at negative flux.}
\label{fig:deboost}
\end{center}
\end{figure}

To search for a difference between the actual and assumed profiles, we examine the profiles of the twenty most significant sources in the ACT map. We fit the data with a background term plus an altered ACT beam (respectively broadened or squeezed by the convolution or deconvolution with a Gaussian). The Gaussian convolution we employed models the broadening of the beam resulting from the random pointing error. The average best-fitted profile is the beam from \citet{hincks/etal:prep} convolved with a Gaussian with $\sigma=+3'' \pm 1''$. This smearing is consistent with the pointing rms in Figure $\ref{fig:pointing}$. The fit is performed out to a radius of 3$'$.  Figure \ref{fig:beam} shows the fit for the brightest source. The effect of this convolution on the 84$''$ full width at half maximum (FWHM) of the ACT beam is at the sub-percent level: $0.05''$. Flux density misestimation will scale roughly as the ratio of the actual profile solid angle to the assumed profile solid angle. In the present case, the misestimation is below 1\%.

As an additional check on the reported flux densities, the matched filter based flux density estimates are compared to $1'$-diameter aperture flux densities. The aperture flux density is defined as the background subtracted flux within a circle (the aperture) centered on the source. The aperture flux density is estimated from the unfiltered ACT maps and as such is complementary to the flux density estimates from the filtered maps.  As the source profile extends well beyond a $1'$ diameter, the aperture flux density estimates are expected to be biased low with respect to true flux density. For the thirty brightest sources from the ACT data, the average ratio of aperture flux density to matched filter flux density is $0.610 \pm 0.003$. In simulations which featured sources with the ACT beam shape, the same ratio is $0.607 \pm 0.004$. If the source profiles in the data were different from the ACT beam shape, then this ratio from the simulation would differ from the data. The agreement between data and simulation further bolsters the claim that flux density bias due to source profile misestimation is at the sub-percent level.

\subsubsection{Deboosting}
\label{subsec:deboosting}

In a source population for which the counts are a steeply falling function of flux, a source's measured flux density ${ S_m}$ in the presence of noise is likely to be an overestimate of its intrinsic value ${ S_0}$. The overestimation is most pronounced for sources at low S/N. The process of deboosting accounts for this overestimate by constructing the posterior flux distribution based on prior knowledge of the source population. Given the relatively high flux density and significance of the detections in this work, we adopt the straightforward Bayesian approach from \cite{coppin/etal:2005}:  
\begin{equation}
{ P(S_0|S_m)} \propto { P(S_m \mid S_0) P(S_0)}
\end{equation}
where the probability ${P(S_m\mid S_0)}$ of measuring ${S_m}$ given ${\rm S_0}$ is taken to be normal with mean ${S_0}$ and  variance derived from the S/N. 

The prior probability of flux $S_0$ in a pixel, ${P(S_0)}$, is computed by generating simulations of the (filtered) intrinsic source flux distribution per pixel. Individual, filtered source profiles, $T_0 \int \Phi(\vec{k}) \tilde{b}(\vec{k}) {\rm d}\vec{k}$, are added to a blank survey map at randomly chosen locations. The numbers and associated amplitudes, $T_0$, of the sources are chosen in accordance  with infrared and radio counts from \citet{toffolatti/etal:1998} rescaled to fit the counts from this study (See Section \ref{subsec:counts}.). The procedure for computing the prior probability based on source counts fit to our data involved an initial rescaling of the \citet{toffolatti/etal:1998} data model to counts computed with raw flux densities. Given deboosted flux densities based on this initial prior, counts were then recomputed, the prior was re-estimated and and used to obtain new deboosted flux densities. This process was iterated until corrections to the counts and the deboosted flux densities became negligible. Only one iteration was required for the relatively small level of deboosting used in this initial study.  Furthermore, we cut off the radio counts at 150 mJy, reflecting the fact that the brightest six sources (${\rm S} > 150$~mJy) are detected and subtracted before constructing the rest of the catalog. A function $P_i(S_0)$ is then generated by binning the fluxes associated with map (indexed here by $i$) pixels in 0.5 mJy bins. The final distribution ${P(S_0)}$ was then computed as the average of  $P_i(S_0)$ from ten-thousand independent simulations. The deboosting algorithm is illustrated for a single source in Figure \ref{fig:deboost} where the dashed Gaussian represents ${P(S_m\mid S_0)}$ and the dotted profile peaking just below zero flux is the prior probability ${P(S_0)}$. The posterior probability ${P(S_0|S_m)}$ is the solid line. 

The deboosted flux ${S_{\rm db}}$ reported in the ACT catalog for sources below~50 mJy is the median of the associated ${P(S_0|S_m)}$, and the reported asymmetric errors enclose the 68\% confidence interval. The abrupt 150 mJy cutoff imposed on the radio counts in combination with finite pixel size effects the smoothness of the prior estimate ${P(S_0)}$ at higher fluxes. To mitigate this effect, the simulated maps used to construct the prior feature a pixel size of half that of the data ($0.25\arcmin$). Furthermore, deboosted fluxes are only provided for sources with flux below 50 mJy where the computed prior is smooth: for sources above 50~mJy, we simply report the center and 68\% confidence level of $P(S_m|S_0)$.

The prior probability ${P(S_0)}$ in Figure \ref{fig:deboost} is broader and, for the range of flux densities plotted, more symmetric than analogous distributions derived in previous work (e.g., Figure 6 of \cite{scott/etal:2008}). The difference arises because our simulations include the radio population whose source counts are much shallower than the infrared populations. The  bright radio sources, in addition to having a bright positive tail, produce significant negative ringing when filtered. A more familiar form of  ${P(S_0)}$ would be generated if we were to recover fluxes using the CLEAN technique \citep{hogbom:1974} as described in, e.g., \cite{vieira/etal:2010}. 

In future multi-band work that considers detections of lower significance, we will employ important extensions of the simple deboosting used here, e.g., \cite{austermann/etal:2010,crawford/etal:inprep}.

\begin{figure}[ht]
\begin{center}
   \plotone{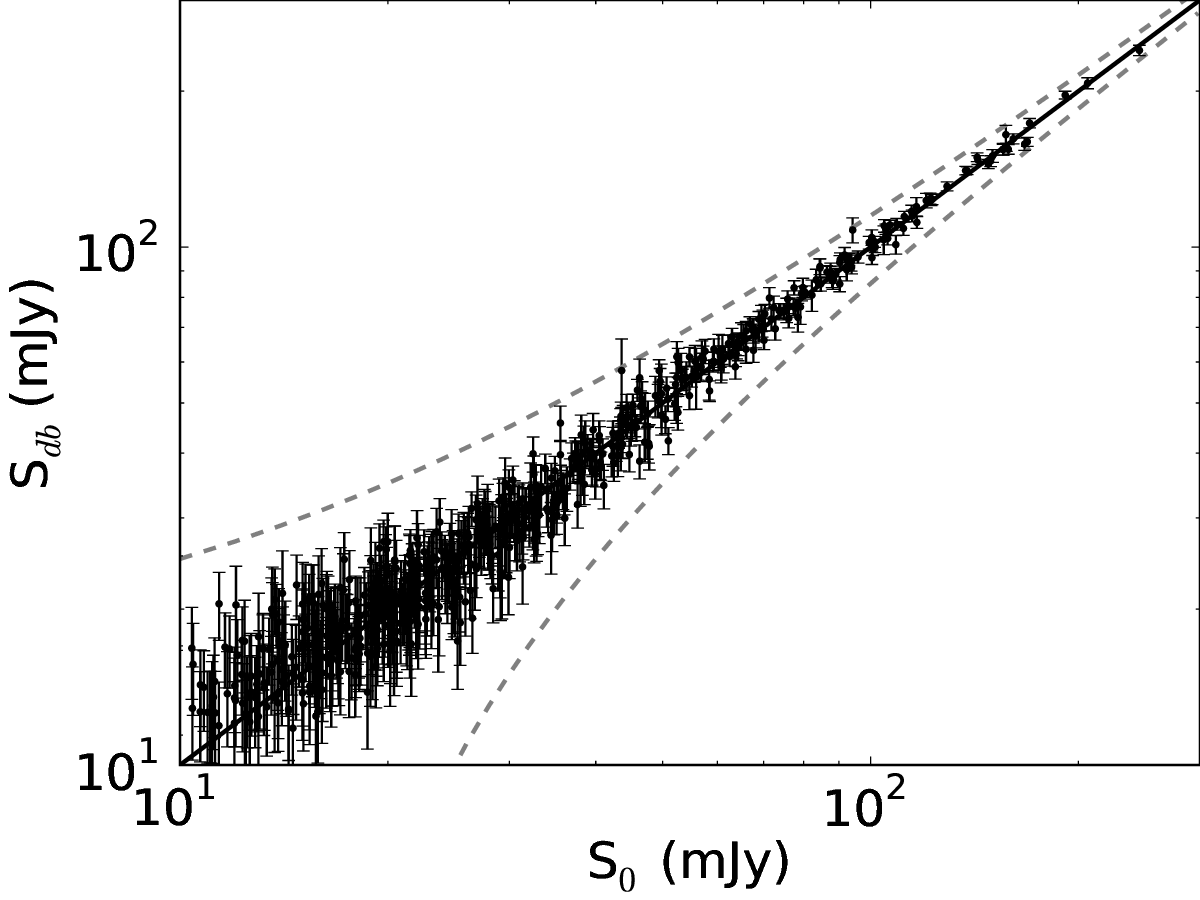}
 \caption{Simulation Flux Recovery. The match-filter derived and deboosted flux density estimate S$_{\rm db}$ plotted versus the intrinsic flux density. The dashed gray lines show the $\pm15$\,mJy limits about the solid black one-to-one line. For S$_{\rm 0}$ $<$ 20\,mJy, the sample is incomplete due to the S/N $>$ \significance\ selection. This is manifest in the graph by the apparent skew of the population above the one-to-one line for low flux densities.  }
\label{fig:recoveredFlux}
\end{center}
\end{figure}

\subsubsection{Simulations and Flux Recovery}
\label{subsec:SimulationsAndFluxRecovery}
Having established that the source profile used in the matched filter is a good approximation to that in the data  (Section \ref{subsec:beam}), it remains to test the flux estimation and deboosting through simulation. Source flux densities were derived from maps with celestial components and noise modeled as described in Section \ref{subsec:dataModeling} with two exceptions. First,  the SZ component was excluded to prevent confusion due to cluster-galaxy correlations. This effect caused negative outliers in derived flux density due to brightness cancellation by the cluster decrement. Second, we excluded several ($>10$) peculiar instances of superposed, very bright, (30 mJy) infrared sources. This apparent clustering of ultra-bright dusty sources caused positive outliers in derived flux density. 

With the exceptions noted above, the simulated sources were detected with the same blind algorithm that was described for source extraction from the data. Similarly, the deboosted flux densities of the simulated sources were derived.  Figure \ref{fig:recoveredFlux} shows the result. At the percent level, the derived deboosted flux density S$_{\rm db}$ is a consistent estimate of the intrinsic flux density S$_{\rm 0}$ with a best fit of
\begin{equation}
{\rm S}_{\rm db} = -0.09\pm0.29 {\rm mJy} + 1.008\pm 0.005 \times {\rm S}_{0}
\end{equation}
where the fit (reduced $\chi^2 = 1.06$ for 451 degrees of freedom) was performed only for sources with flux density greater than 20\,mJy to restrict the analysis to a complete (and thus symmetric) distribution. The errors (68\% c.l.) are derived from one thousand bootstrap samplings. Furthermore, the model S$_{\rm db}$ = S$_{\rm 0}$ fits the 451 sources with flux densities greater than 20\,mJy with a reduced  $\chi^2 = 1.07$. This one-to-one model fits the entire population  (down to the S/N=\significance) with reduced $\chi^2$=1.19 over 651 degrees of freedom.  This improbable statistic likely results from a combination of unaccounted-for astrophysics (e.g., source clustering) in \cite{sehgal/etal:2010}, errors in our deboosting, and an underestimate of flux density error. 

\subsubsection{Convergence}
\label{subsec:convergence}
As described in Section \ref{sec:data}, maximum-likelihood maps generated from the cross-linked ACT \arone\ data are unbiased for modes corresponding to multipoles in excess of a few hundred (i.e., exactly those used for source flux estimation).  We checked this claim through signal-only simulations and found that the flux density estimates of the sources converged to within 1\% of their simulation values before the twenty-fifth iteration of the preconditioned conjugate gradient map-solver (See \citet{fowler/etal:2010} for a fuller description of the solver.). This study uses a map from the hundredth iteration of the ACT maps. Therefore, we conclude that the flux densities are not biased by failure of the mapping algorithm convergence.

\begin{deluxetable}{ccccc}
\tablecaption{Number Counts, Purity, and Completeness.\tablenotemark{a}}
\tablehead{
\colhead{Flux Range} & \colhead{\# Det} & \colhead{\# False} & \colhead{Completeness} & \colhead{Area\tablenotemark{b}} \\
\colhead{mJy} &  & &  & \colhead{sq-deg} 
}
\startdata
1390 -- 2870	& 1		& 0		& 100\% 		& 432 \\
\,\,\,650 -- 1390	& 1		& 0		& 100\%		& 432 \\
330 -- 650	& 2		& 0		& 100\%		& 432 \\
170 -- 330        & 4           & 0           & 100\%           &  432 \\
\,\,\,90 -- 170	& 12		& 0		& 100\% 		& 432 \\
50 -- 90		&  22	& 0		& $94\pm6$\% 	& 432 \\
30 -- 50		&  40	& 0		& 100\% 		& 368 \\
20 -- 30		&  44	&$1\pm1$	& $97\pm2$\% & 368 \\
15 -- 20		&  25	&$2\pm1$& $86\pm5$\% & 368 \\
\enddata
\tablenotetext{a}{See Figure \ref{fig:dNdS} for a graph of purity/completeness-corrected differential source counts.}
\tablenotetext{b}{For a discussion of flux density dependent area, see Section \ref{subsec:sourceDetection}.}
\label{tab:stats}
\end{deluxetable}

\subsection{Purity and Completeness}
\label{subsec:purityAndCompleteness}
The number of false detections in our catalog of \numberOfSources\ sources is estimated by running the detection algorithm on an inverted (negative temperature) map in which the SZ decrements from all ACT-detected and optically-confirmed clusters have been masked. A full description of the ACT SZ cluster population and optical follow-up can be found in \citet{menanteau/etal:2010}. With this approach, three spurious detections are found, giving a purity of 98\% for detections above a S/N of \significance. Below this S/N, the purity of the sample was found to decrease rapidly with only $\sim$50\% purity in the range $5 < {\rm S/N} < \significance$ (seven false detections).  These results are consistent with estimates of purity based on cross-identification of the ACT detections with other catalogs (See Section \ref{subsec:others}.). From simple Gaussian statistics, one expects fewer than five false detections in a sample selected with S/N $>$ 5. Thus, some fraction of the false detections are may be the result of localized noise not accounted for by the weighting and matched filter.

Simulations from \citet{sehgal/etal:2010} with noise from difference maps (see Section \ref{subsec:dataModeling}) were used to estimate completeness. Table \ref{tab:stats} summarizes the findings. Due primarily to the uneven depth of coverage, the population of sources detected between 15 and 20 mJy was found on average to be 86\% complete and the population between 20 and 30 mJy to be 97\% complete. Because of the strict 50\,mJy lower bound set for detections in areas of the map characterized by integration times below 7 minutes, the 50 -- 90\,mJy range also suffers from an incompleteness of 94\%. The full simulations of \citet{sehgal/etal:2010} include  a correlation between radio sources and galaxy clusters, and we have excluded the SZ component in order to simplify the current study. At \arone, clusters manifest themselves as arcminute-scale temperature decrements in the map which will cancel source flux in superposed source-cluster pairs. We also ran the test described here with the SZ component from \citet{sehgal/etal:2010}  included in the simulations. The resulting cancellation of source flux density by cluster decrements was found to cause an additional few percent of the incompleteness in the source population with flux densities below 30 mJy.

\section{Discussion}
\label{sec:discussion}

\subsection{Comparison to Other Source Catalogs}
\label{subsec:others}
As a first step in ACT source characterization we consider cross-identifications with other catalogs. Matches are established within a 30$\arcsec$ radius about an ACT source. The choice of association radius was made based on the positional rms of the ACT catalog and comparison catalogs (allowing for outliers) as well as the fact that the source of low-frequency radio signals in a given system may be physically displaced from the source of high-frequency radio signal. A general search through the NASA/IPAC Extragalactic Database finds that thirty-one ACT sources have measured redshifts, ranging from 0.003 to 2.46. In what follows we consider in more detail several catalogs of particular relevance to the \arone\ source population.

Of our \numberOfSources\ sources, 109 match sources in the 5, 8, and 20\,GHz AT20G catalog.   There are 180 AT20G sources in the survey area such that a random cross-identification  would occur once in roughly 11600 cases.\footnote{This rough statistic estimates the probability of a spurious detection in the ACT data falling in the fractional area  ($N \times \pi \times 30\arcsec^2$) occupied by the $N$ sources from the auxiliary catalog. } The AT20G catalog is incomplete below 100 mJy \citep{murphy/etal:2010}. Given that nearly all the radio sources detected at \arone\ are expected to have relatively flat spectra \citep[e.g.,][]{vieira/etal:2010}, faint ACT sources may not have matches in AT20G (See Section \ref{subsec:spectra}.). We have proposed for time on the Australia Telescope Compact Array to measure flux densities for sources in the ACT catalog that do not appear in AT20G.

All but six of the ACT sources are co-located within 30$''$  of sources from AT20G or the 0.84\,GHz Sydney University Molonglo Sky Survey (SUMSS) catalog \citep{mauch/etal:2003}. Within our survey area, the sample of 14030 SUMSS sources is complete to 8 mJy. A random cross-identification with a SUMSS detection is a 1-in-150 event and thus a spurious SUMSS association is likely. Of the six ACT sources without cross-identification in AT20G or SUMSS, two (ACT-S J011830$-$511521, ACT-S J033133$-$515349) are within 50$''$ of a SUMSS source, and the former is relatively bright with a 148 GHz flux density of 47.6 mJy. Furthermore, a preliminary reduction of the ACT \artwo\ data identifies one of the remaining four (ACT-S J031823$-$533148) as a 5$\sigma$ detection.  ACT-S J034157$-$515140, ACT-S J004042$-$511830, and ACT-S J035343$-$534553 have no match in the auxiliary catalogs and may be false detections. This number of false detections is consistent with the study of sample purity presented in Section \ref{subsec:purityAndCompleteness}.

Comparing to the recently reported 2.0 mm measurements from the South Pole Telescope (SPT) \citep{vieira/etal:2010}, we find twenty-four cross-identifications with ACT sources. The \citet{vieira/etal:2010} study used a square survey of 87 square-degrees centered at 05h right ascension. As such, the ACT and SPT surveys have only fractional overlap. Nevertheless, 2304 of the 3496 SPT source candidates (S/N $>3$) fall within the ACT survey. All twenty-four
 matching sources were categorized in \citet{vieira/etal:2010} as synchrotron-dominated.

Finally, we compare the ACT catalog to the Infrared  Astronomical Satellite Point Source Catalog \citep[IRAS PSC;][]{helou/walker:1998}. Three of the detections, ACT-S J041959$-$545622 (NGC 1566), ACT-S J04285$-$542959 and ACT-S J033133$-$515352 (IC 1954), are identified with sources in the IRAS PSC. All three sources display lower frequency radio emission and have been identified in the preliminary ACT \artwo\ analysis.

\subsection{Correlation with X-ray Clusters}

Radio-loud AGN are frequently found in Brightest Cluster Galaxies (BCGs). From a study of radio-loud AGNs in
the SDSS using data from the National Radio Astronomy Observatory (NRAO) Very Large Array (VLA) Sky Survey
\citep{condon/etal:1988} and the Faint Images of the Radio Sky At Twenty Centimeters survey 
\citep{becker/etal:1995}, \citet{best/etal:2007} found that the probability of a BCG hosting a radio-loud AGN is significantly
enhanced compared to field galaxies of the same stellar mass. \citet{lin/mohr:2007} found that the 
fraction of BCGs being radio-loud is higher compared to that of cluster galaxies of similar luminosity.
This enhancement of radio activity is probably due to the fact that
BCGs are located in special places - the centers of clusters - and the
AGN activity is likely fueled by gas cooling, or due to galaxy
interactions in these high density regions.

We performed a simple check to test whether any of the ACT \arone\ sources were associated with clusters by
cross matching against the REFLEX catalog \citep{bohringer/etal:2004}, which is a homogeneous, X-ray selected
sample with a nominal (0.1-2.4~keV) flux limit of $\gtrsim 3 \times 10^{-12}$~erg~s$^{-1}$ covering $\delta <
+2.5\deg$. The REFLEX catalog is $> 90$\% complete above its nominal flux limit, and 23 REFLEX clusters are
located within 2008 ACT survey region, spanning the redshift range $0.03 < z < 0.34$. Using a
generous $5\arcmin$ matching radius, we find that three ACT sources are associated with REFLEX clusters:
ACT-S J062142-524136 (RXC J0621.7-5242); ACT-S J042906-534943 (RXC J0429.1-5350/AS0463); and ACT-S
J062620-534136 (RXC J0626.3-5341/A3391). The separation between the ACT sources and the corresponding REFLEX
cluster positions is $0.3-1.1\arcmin$. All of these clusters are at very low redshift
(0.041--0.055), and have low masses ($\sim (0.4-2) \times 10^{14}$~M$_\sun$, inferred from their X-ray
luminosities).

All three of these ACT sources have corresponding matches in the AT20G catalog. For ACT-S J062142-524136, both the
ACT and AT20G sources are located within $<5\arcsec$ of each other and are coincident with the BCG, from
inspection of DSS imaging. For the other two sources, we find that either the ACT source position (in the
case of RXC J0626.3-5341) or the AT20G position (in the case of RXC J0429.1-5350) is coincident with the BCG,
although the AT20G and ACT positions are offset by $< 19\arcsec$. In all cases, the projected radial distance
between the BCG and the REFLEX X-ray position is $< 60$~kpc.

\begin{figure}
\begin{center}
    \plotone{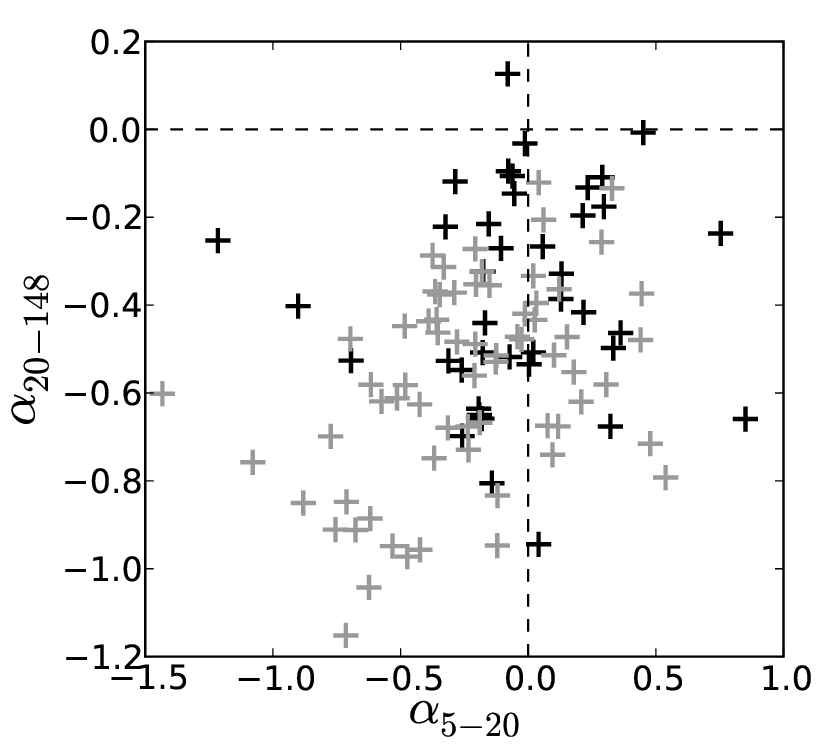}
 \caption{Radio source spectral indices. In this radio color-color diagram, 5--20\,GHz and 20--\arone\  spectral indices are shown for ACT-AT20G cross-identified sources. The population is dominated by sources which are peaked (lower right quadrant) or falling (lower left quadrant). Black (Gray) crosses correspond to sources with \arone\ flux density greater (less) than 50 mJy. The low flux sample is incomplete and suffers from selection bias that favors sources with more negative spectral indices.}
\label{fig:spectralIndices}
\end{center}
\end{figure}

\subsection{Source Spectra}
\label{subsec:spectra}
It is established that radio source spectra extending to \arone\ are not well characterized by a simple power law S($\nu$) $\propto \nu^\alpha$ (e.g., \citet{dezotti/etal:2010}). \citet{murphy/etal:2010} used a color-color comparison of spectral indices at 5--8\,GHz and 8--20\,GHz to show that the AT20G population may be decomposed into classical steep (and steepening) spectrum sources, sources that peak between their bands, and sources that show flat, rising or up-turned spectra. Following this example we construct a color-color comparison of 5--20\,GHz and 20--\arone\ measurements, where the 5 and 20\,GHz flux densities are from AT20G. The variability of these sources makes a per-source comparison difficult, but a study using the 109 ACT-AT20G cross-identifications as an ensemble is meaningful. Figure \ref{fig:spectralIndices} is the $\alpha_{5-20}$ vs. $\alpha_{20-148}$  color-color diagram. The figure shows that the ACT-AT20G cross-identified sources are predominantly characterized by spectral steepening. The sample is biased towards steepened spectra at low flux density ($S_{148} < 50$~mJy) due to incompleteness in AT20G. 

The population can be further divided according to \arone\ flux density. In Figure \ref{fig:spectralIndices}, the black crosses correspond to the forty-two brightest sources in the ACT sample, and the gray crosses correspond to the faint half. The dividing flux density, 50 mJy, was chosen such that all but two of the ACT detections in the brighter sample have cross-identifications in AT20G. As described in Section \ref{subsec:others}, below this flux density the mean spectral indices of the population of ACT-AT20G cross-identified sources is biased negative by the incompleteness in AT20G below 100 mJy. The two sources in this high-flux subset which do not have AT20G counterparts are characterized by 150 GHz flux densities close to the 50 mJy cutoff and may have flatter spectra than the sample average. Including two extra sources with $\alpha=0$ changes the mean spectral indices by $\sim 25\%$ of the statistical error, and therefore we simply use the forty sources with AT20G counterparts in the following analysis. Considering only the unbiased, bright half of the distribution, the average spectrum steepens between 5--20\,GHz and 20--\arone. The median spectral indices of the unbiased sample of sources are $\alpha_{5-20} = -0.07 \pm 0.37 (\pm 0.06)$ and $\alpha_{20-148} = -0.39 \pm 0.24 (\pm 0.04)$.\footnote{The errors on spectral indices are the 68\% confidence levels of the distribution and, in parentheses, for the median.} In obtaining the spectral indices we compare the deboosted flux densities from ACT with the raw flux densities from AT20G.  Using the raw AT20G should not significantly bias the index estimates because the AT20G detections are all characterized by a S/N greater than 15.

\begin{figure}[ht]
\begin{center}
   \plotone{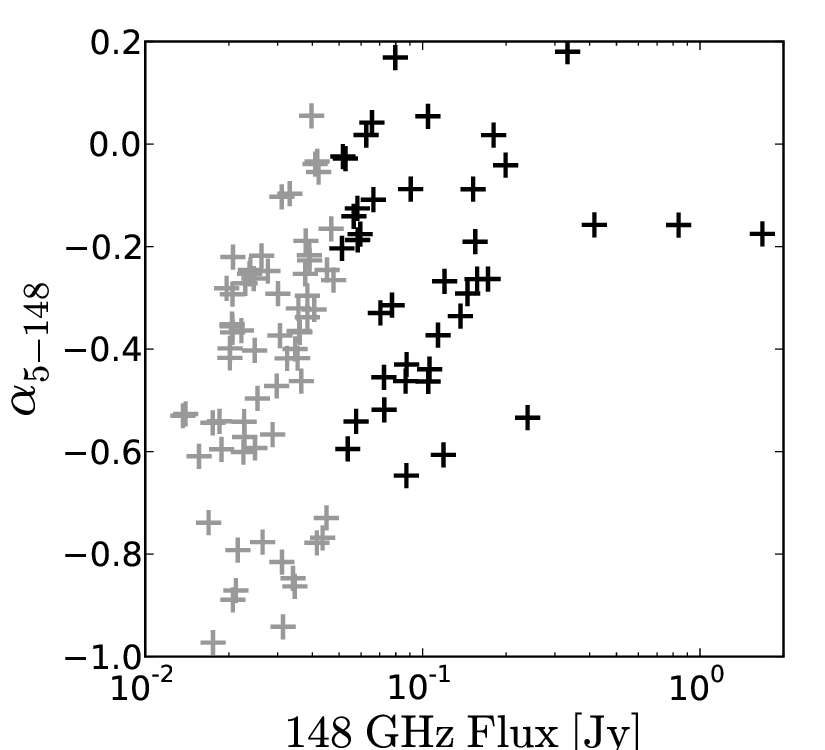}
 \caption{Radio color-magnitude diagram. As in Figure \ref{fig:spectralIndices}, the data are divided between high flux densities (black crosses) and low flux densities (gray crosses) at 50 mJy. The unbiased 50 mJy sample has a 5--\arone\ spectral index of $\alpha_{5-148} = -0.20 \pm 0.21 (\pm 0.03)$. The low flux density data suffer a selection bias that excludes sources with flat or rising spectra.}
\label{fig:colormag}
\end{center}
\end{figure}

When restricted to the unbiased sample with flux densities above 50 mJy, the 5--\arone\ spectral index is $\alpha_{5-148} = -0.20 \pm 0.21 (\pm 0.03)$.  This distribution is in 2.5 $\sigma$ tension with the SPT-reported mean 5--150\,GHz spectral slope (for 57 sources) of $\alpha_{5-150} = -0.13 \pm 0.21 (\pm 0.03)$ \citep{vieira/etal:2010}.\footnote{Note the tension arises for the \emph{center} of the spectral index distribution which is better constrained than the index of any given source.}  \citet{vieira/etal:2010}  claim that the mean spectral index of the synchrotron-dominated species remains near $-0.1$ to 2.0 mm ($\approx148$~GHz) after which it steepens such that the average slope between 2 mm and 1.4 mm ($\approx220$~GHz) is $-0.5$. This study suggests that the transition to the steep spectrum is more gradual. In fact, the spectral slope $\alpha_{20-148} = -0.39 \pm 0.24 (\pm 0.04)$ approaches the $-0.5$ slope between 2.0 mm and 1.4 mm reported in \cite{vieira/etal:2010}. This picture is further supported by the rescaling of ACT \arone\ source counts relative to source counts at 30 GHz (See Section \ref{subsec:counts}). Figure \ref{fig:colormag} shows the ACT 5--\arone\  spectral indices as a function of flux density. The low flux density sample, represented by gray crosses, is incomplete for high spectral indices. Follow-up of the 148~GHz selected sources without matches in AT20G will complete the picture in the range 20--50 mJy.

\begin{figure*}[ht]
\begin{center}
  \resizebox{6.5in}{!}{
     \plotone{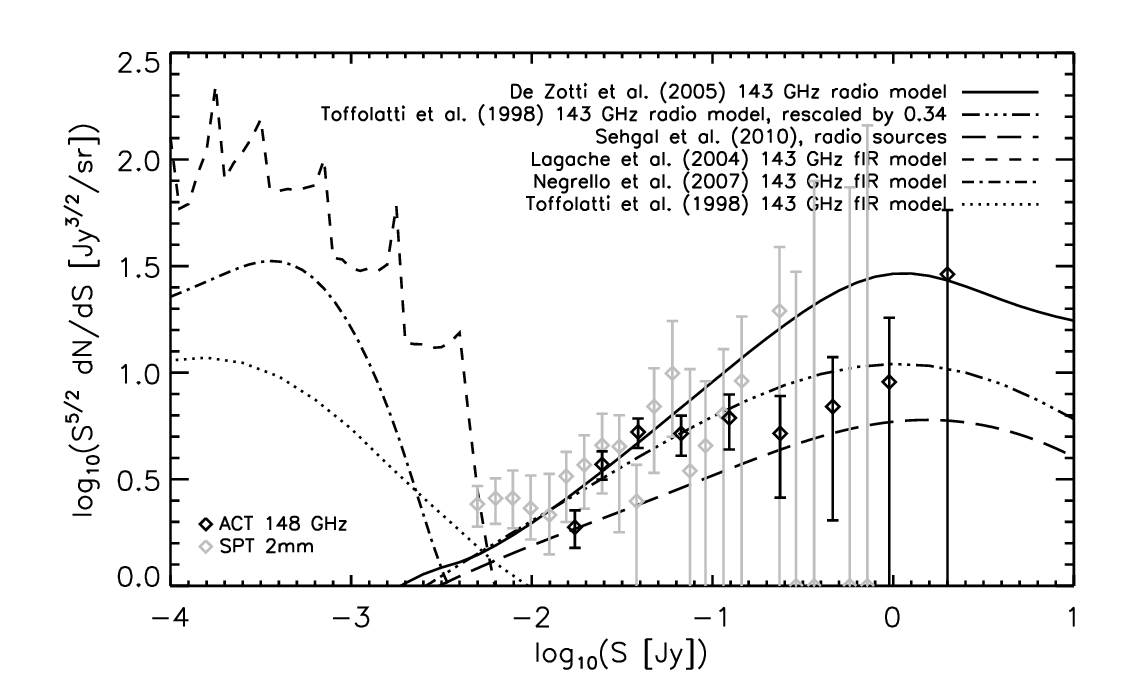}
  }
\caption{ \arone\ Differential Source Counts. Derived from Table \ref{tab:stats} and corrected for completeness, the ACT differential source counts (bold diamonds) are plotted together with models of radio and infrared source populations at 143\,GHz (as computed for the 2 mm band of the Planck satellite). The data are consistent with being dominated by radio sources. Data (gray diamonds) from the SPT \citep{vieira/etal:2010} are consistent with the ACT counts. Errors (1$\sigma$) are Poissonian. }
\label{fig:dNdS}
\end{center}
\end{figure*}

\subsection{Source Counts}
\label{subsec:counts}
The differential number counts for ACT sources based on data in Table \ref{tab:stats} are plotted in Figure \ref{fig:dNdS}. The figure shows that the ACT counts are fit reasonably well by the model for radio sources from \citet{toffolatti/etal:1998} scaled by a factor of $0.34\pm0.04$. Depending on whether the rescaling is fit to the data in log or linear coordinates the best fit value varies from 0.31 to 0.35. As in Figure \ref{fig:dNdS}, we adopt the rescaling of 0.34 which lies between these two values. The ACT results are also consistent with counts reported by the SPT \citep{vieira/etal:2010}. As part of the WMAP 7-year analysis,  \cite{wright/etal:2009} fit differential source counts at 30\,GHz from WMAP, the VSA \citep{cleary/etal:2006}, CBI \citep{mason/etal:2003}, and DASI \citep{kovac/etal:2002}.  They found that the best fit scaling for the \citet{toffolatti/etal:1998} model  at 30\,GHz is 0.64. This result was also consistent with 30~GHz measurements by the SZA which covered a fainter range of flux densities \citep{muchovej/etal:2010}. The apparent discrepancy between the 30\,GHz and \arone\ scalings can be explained by the steepening of the radio source spectrum described in Section \ref{subsec:spectra}. The source count model traditionally adopted from  \citet{toffolatti/etal:1998} uses an average spectral index of $\alpha=0$ for radio sources between 20 and 200\,GHz. From the unbiased sample of spectral indices (dark crosses) in Figure \ref{fig:spectralIndices}, the slope between 20 and \arone\ is $\alpha \approx -0.4$. As can be seen from Figure \ref{fig:dNdS}, the radio source number counts are well-approximated by a power law N($>$S) $\propto$ S$^{-1}$ between 0.01 and 1 Jy (dN/dS S$^{5/2}$ $\propto$ S$^{1/2}$). It follows that the \arone\  number counts should be rescaled by 
\begin{eqnarray}
0.64 \times \frac{S_{\rm\arone}}{S_{\rm30\,GHz}} & \approx & 0.64 \times \left(\frac{\rm \arone}{\rm 30\,GHz}\right)^\alpha \\ \nonumber
& \approx&  0.64 \times 5^{-0.4}\\ \nonumber 
& \approx& 0.35, \nonumber
\end{eqnarray}
consistent with the scaling of \citet{toffolatti/etal:1998} to the ACT data. Also shown in the plot are radio source models from \cite{dezotti/etal:2005} and \cite{sehgal/etal:2010}. The \cite{dezotti/etal:2005} model is consistent with the ACT source counts at low flux densities ($S<0.1$ Jy) and over-predicts the counts at higher flux densities.   The model presented in \cite{sehgal/etal:2010}, although underestimating the source counts at $S<0.1$ Jy, seems to be consistent with ACT data at higher fluxes.

Differential source counts derived from the 150\,GHz catalog from ACBAR \citep{reichardt/etal:2009a} are consistent with \citet{toffolatti/etal:1998} scaled by 0.64. However, ACBAR's resolution and sensitivity made it sensitive to only the brightest sources with flux densities mainly in excess of 100 mJy where there are fewer sources. Furthermore,  the ACBAR fields were chosen specifically to have bright quasars for beam measurement \citep{kuo/etal:2007} such that the differential source counts derived from the ACBAR data are biased high. The ACT field was chosen with no a-priori knowledge of source population. Allowing for variability, there is reasonable agreement in the flux densities reported by ACBAR and ACT for the seven sources common to both catalogs.

Also shown in Figure \ref{fig:dNdS} are models for counts of dusty starburst galaxies \citep{toffolatti/etal:1998,lagache/etal:2004,negrello/etal:2007}. The brightest infrared sources in these models are 10 mJy. Given that all sources in the ACT catalog have flux densities greater than 10 mJy, these models predict that the \arone\  selected ACT catalog should have few or no infrared sources.

\subsection{Contribution to the Power Spectrum}
\label{subsec:powerSpectrum}
The $\ell$=3000--10000 power spectrum of the \arone\ sky is dominated by synchrotron and infrared sources as well as the thermal SZ from clusters \citep{hall/etal:2010,fowler/etal:2010} . In particular, the power spectrum at the highest multipoles constrains the Poisson distributed component of the source population, and  an understanding of the residual synchrotron population helps break the high-$\ell$ spectral degeneracy between synchrotron and  infrared sources. The contribution to the power spectrum by a Poisson-distributed  population of sources is a function of the number counts:
\begin{equation}
C^{\rm PS} = \left(\frac{\delta B_\nu}{ \delta T}\right)^{-2} \int_0^{S_{\rm lim}}  S^2 \frac{{\rm d}N}{{\rm d}S} {\rm d}S,
\label{eqn:cl}
\end{equation}
where $S_{\rm lim}$ is the upper-limiting flux density of the residual (i.e., unmasked) sources in the data. With a limiting flux density of 20~mJy and the rescaled model of \cite{toffolatti/etal:1998} from Section \ref{subsec:counts}, one expects a synchrotron contribution to the Poisson power spectrum of $C^{\rm Sync} = (2.8 \pm 0.3) \times 10^{-6} \micro{\rm K}^2$.

\citet{fowler/etal:2010} used a 20~mJy cut for masking sources and found a Poisson spectrum from all residual sources of $C^{\rm PS} = (7.8 \pm 2.3) \times 10^{-6} \micro{\rm K}^2$. Thus from this study we expect an infrared contribution to the Poisson spectrum of $C^{\rm IR} = (5.0 \pm 2.3) \times 10^{-6} \micro{\rm K}^2$. A similar argument applied to the study of \citet{lueker/etal:2010} results in an estimate of the residual Poisson term for infrared sources of $C^{\rm IR} \approx (6.3 \pm 0.5) \times 10^{-6} \micro{\rm K}^2$. Thus, within the  \citet{fowler/etal:2010} errors, the two studies of the high-$\ell$ power spectrum at \arone\ are consistent.

\section{Conclusions}

We have presented results on extragalactic sources at \arone\ from data taken by ACT during the 2008 observing season. A catalog of \numberOfSources\ millimeter sources has been presented with sources detected across two decades in flux density, from \minimumFlux\ to \maximumFlux. The flux density calibration of the sources derives from observations of Uranus with 6\% error. Bias in the quoted flux densities due to beam shape uncertainty is estimated at less than 1\%. Typical statistical 1$\sigma$ errors for the source flux  density range from 2.5 mJy to 5 mJy. The catalog astrometry error for the brightest sources is characterized by an rms of $3.5\arcsec$. The catalog is estimated to be 98\% pure and complete above 20 mJy. 

Comparison to other catalogs shows that 98\% of the ACT detections correspond to sources detected at lower radio frequencies.  The differential source counts are also consistent with the finding that ACT detections correspond to sources detected at lower radio frequencies. In particular, the source counts are fit reasonably well by the radio model of \cite{toffolatti/etal:1998} scaled by 0.34.  This scaling, compared to a scaling 0.64 found at 30\,GHz by \citet{wright/etal:2009}, suggests that the population of radio sources is characterized, on average, by spectral steepening between 30\,GHz and \arone. This conclusion is consistent with the average spectral indices derived from the combined AT20G and ACT datasets. Future work will address the more involved comparison with the \cite{dezotti/etal:2005} and \cite{sehgal/etal:2010} radio source models. With the rescaled  model from  \cite{toffolatti/etal:1998} and a 20~mJy cut, the residual contribution of the synchrotron population to the Poisson power spectrum is $C^{\rm Sync} = (2.8 \pm 0.3) \times 10^{-6} \micro{\rm K}^2$.

 Future ACT source work will incorporate the \artwo\ and \arthree\ bands, deeper coverage integrating the 2007, 2009, and (ongoing) 2010 seasons, as well as the equatorial survey overlapping the deep Sloan Digital Sky Survey Stripe 82.

\acknowledgments
The ACT project was proposed in 2000 and funded by the U.S. National Science Foundation on January 1, 2004. Many
have contributed to the project since its inception. We especially
wish to thank Asad Aboobaker, Christine Allen, Dominic Benford, Paul
Bode, Kristen Burgess, Angelica de Oliveira-Costa, Peter Hargrave,
Norm Jarosik, Amber Miller, Carl Reintsema, Felipe Rojas, Uros Seljak, Martin
Spergel, Johannes Staghun, Carl Stahle, Max Tegmark, Masao Uehara,
Katerina Visnjic, and Ed Wishnow. It is a pleasure to acknowledge Bob
Margolis, ACT's project manager. Reed Plimpton and David Jacobson
worked at the telescope during the 2008 season. ACT is on the
Chajnantor Science preserve, which was made possible by the Chilean
Comisi\'on Nacional de Investigaci\'on Cient\'ifica y Tecnol\'ogica.

This work was supported by the U.S. National Science Foundation
through awards AST-0408698 for the ACT project, and PHY-0355328,
AST-0707731 and PIRE-0507768. Funding was also provided by Princeton
University and the University of Pennsylvania.  The PIRE program made
possible exchanges between Chile, South Africa, Spain and the US that
enabled this research program.  Computations were performed on the GPC
supercomputer at the SciNet HPC Consortium.  SciNet is funded by: the
Canada Foundation for Innovation under the auspices of Compute Canada;
the Government of Ontario; Ontario Research Fund -- Research
Excellence; and the University of Toronto.

TM was supported through NASA grant NNX08AH30G. JBJ was supported by the 
FONDECYT grant 3085031. ADH received additional support from a Natural 
Science and Engineering Research Council of Canada (NSERC) PGS-D scholarship. 
AK and BP were partially supported through NSF AST-0546035 and AST-0606975,
respectively, for work on ACT\@.  HQ and LI acknowledge partial support
from FONDAP Centro de Astrof\'isica. NS is supported by the U.S. Department of Energy contract 
to SLAC no. DE-AC3-76SF00515. RD was supported by CONICYT,
MECESUP, and Fundaci\'on Andes. RH was supported by the Rhodes Trust.  
ES acknowledges support by NSF Physics Frontier Center grant PHY-0114422 to the Kavli Institute of
Cosmological Physics. YTL acknowledges support from the World Premier
International Research Center Initiative, MEXT, Japan.  The ACT data will
be made public through LAMBDA (http://lambda.gsfc.nasa.gov/) and the
ACT website (http://www.physics.princeton.edu/act/).

\appendix



%
%

\clearpage


\LongTables

\renewcommand\thetable{A1}

\begin{deluxetable}{cccccccc}

\tabletypesize{\footnotesize}
\tablecolumns{8} 
\tablewidth{0pt} 
\tablecaption{ACT High Significance \arone\ Extragalactic Source Catalog}
\tablehead{
           \colhead{ACT ID}                                    &
           \multicolumn{2}{c}{RA  (J2000)   Dec}               &
           \colhead{S/N} &
           \colhead{S$_{\rm m}$\tablenotemark{a}}  &
           \colhead{S$_{\rm db}$\tablenotemark{b}}  &
            \colhead{t$_{\rm int}$\tablenotemark{c}}     &
           \colhead{AT20G ID }                 
                                          \\
           \colhead{}                                          &        
           \colhead{\phn{h}\phn{m}\phn{s}}                         &
           \colhead{\phn{\arcdeg}~\phn{\arcmin}~\phn{\arcsec}} &
           \colhead{} &
           \colhead{(mJy)}                                          &            
           \colhead{(mJy)}                                          & 
           \colhead{(min)}               &
           \colhead{}                                          }

\startdata
 ACT-S J001850$-$511454  &  $00:18:50$  &  $-51:14:54$  &  10.4  &  53.7  &  53.7$^{+5.2}_{-5.2}$  &  6.1  & ATG20 J001849$-$511455  \\ 
 ACT-S J001851$-$492949  &  $00:18:51$  &  $-49:29:49$  &  7.4  &  52.8  &  52.8$^{+7.1}_{-7.1}$  &  3.0  & ... \\ 
 ACT-S J002513$-$542737  &  $00:25:13$  &  $-54:27:37$  &  6.7  &  26.1  &  24.8$^{+3.9}_{-4.0}$  &  10.0  & ATG20 J002511$-$542739  \\ 
 ACT-S J003134$-$514308  &  $00:31:34$  &  $-51:43:08$  &  13.9  &  52.7  &  52.7$^{+3.8}_{-3.8}$  &  10.3  & ATG20 J003134$-$514325  \\ 
 ACT-S J003734$-$530729  &  $00:37:34$  &  $-53:07:29$  &  11.9  &  39.1  &  38.5$^{+3.3}_{-3.3}$  &  13.9  & ATG20 J003735$-$530733  \\ 
 ACT-S J004012$-$514950  &  $00:40:12$  &  $-51:49:50$  &  5.7  &  20.2  &  18.7$^{+3.7}_{-3.7}$  &  12.2  & ... \\ 
 ACT-S J004042$-$511830  &  $00:40:42$  &  $-51:18:30$  &  5.6  &  22.9  &  21.1$^{+4.2}_{-4.2}$  &  9.5  & ... \\ 
 ACT-S J004906$-$552106  &  $00:49:06$  &  $-55:21:06$  &  14.1  &  62.6  &  62.6$^{+4.5}_{-4.5}$  &  7.5  & ATG20 J004905$-$552110  \\ 
 ACT-S J004949$-$540240  &  $00:49:49$  &  $-54:02:40$  &  6.7  &  21.8  &  20.7$^{+3.3}_{-3.3}$  &  14.2  & ... \\ 
 ACT-S J005240$-$531127  &  $00:52:40$  &  $-53:11:27$  &  5.5  &  17.6  &  16.2$^{+3.4}_{-3.4}$  &  14.5  & ... \\ 
 ACT-S J005605$-$524154  &  $00:56:05$  &  $-52:41:54$  &  8.2  &  28.6  &  27.6$^{+3.5}_{-3.5}$  &  12.5  & ... \\ 
 ACT-S J005622$-$531845  &  $00:56:22$  &  $-53:18:45$  &  9.7  &  32.4  &  31.6$^{+3.4}_{-3.4}$  &  13.5  & ... \\ 
 ACT-S J005706$-$521423  &  $00:57:06$  &  $-52:14:23$  &  13.9  &  47.4  &  46.8$^{+3.4}_{-3.4}$  &  12.7  & ATG20 J005705$-$521418  \\ 
 ACT-S J005855$-$521925  &  $00:58:55$  &  $-52:19:25$  &  20.3  &  70.4  &  70.4$^{+3.5}_{-3.5}$  &  12.4  & ATG20 J005855$-$521926  \\ 
 ACT-S J010309$-$510907  &  $01:03:09$  &  $-51:09:07$  &  8.8  &  33.5  &  32.5$^{+3.8}_{-3.8}$  &  10.4  & ATG20 J010306$-$510907  \\ 
 ACT-S J010330$-$513544  &  $01:03:30$  &  $-51:35:44$  &  11.7  &  41.9  &  41.2$^{+3.6}_{-3.6}$  &  12.0  & ATG20 J010329$-$513551  \\ 
 ACT-S J011324$-$532938  &  $01:13:24$  &  $-53:29:38$  &  7.2  &  22.4  &  21.4$^{+3.2}_{-3.2}$  &  15.4  & ATG20 J011323$-$532949  \\ 
 ACT-S J011654$-$544653  &  $01:16:54$  &  $-54:46:53$  &  5.8  &  21.0  &  19.5$^{+3.7}_{-3.8}$  &  11.3  & ... \\ 
 ACT-S J011830$-$511521  &  $01:18:30$  &  $-51:15:21$  &  13.0  &  48.4  &  47.7$^{+3.6}_{-3.7}$  &  11.0  & ... \\ 
 ACT-S J011950$-$535714  &  $01:19:50$  &  $-53:57:14$  &  46.9  &  157.2  &  157.2$^{+3.4}_{-3.4}$  &  15.4  & ATG20 J011950$-$535717  \\ 
 ACT-S J012006$-$521104  &  $01:20:06$  &  $-52:11:04$  &  6.1  &  19.6  &  18.4$^{+3.3}_{-3.3}$  &  14.8  & ATG20 J012008$-$521102  \\ 
 ACT-S J012457$-$511309  &  $01:24:57$  &  $-51:13:09$  &  51.0  &  199.1  &  199.1$^{+3.9}_{-3.9}$  &  11.4  & ATG20 J012457$-$511316  \\ 
 ACT-S J012501$-$532514  &  $01:25:01$  &  $-53:25:14$  &  5.3  &  15.5  &  14.2$^{+3.1}_{-3.2}$  &  17.6  & ... \\ 
 ACT-S J012623$-$510305  &  $01:26:23$  &  $-51:03:05$  &  7.0  &  26.0  &  24.8$^{+3.8}_{-3.8}$  &  11.1  & ATG20 J012624$-$510309  \\ 
 ACT-S J012755$-$513642  &  $01:27:55$  &  $-51:36:42$  &  12.1  &  39.5  &  38.8$^{+3.3}_{-3.3}$  &  14.1  & ATG20 J012756$-$513641  \\ 
 ACT-S J012756$-$532933  &  $01:27:56$  &  $-53:29:33$  &  9.6  &  27.6  &  26.9$^{+2.9}_{-2.9}$  &  18.1  & ... \\ 
 ACT-S J012834$-$525520  &  $01:28:34$  &  $-52:55:20$  &  6.3  &  18.5  &  17.4$^{+3.0}_{-3.0}$  &  17.7  & ATG20 J012834$-$525519  \\ 
 ACT-S J013107$-$545707  &  $01:31:07$  &  $-54:57:07$  &  6.3  &  23.8  &  22.3$^{+3.9}_{-3.9}$  &  10.3  & ... \\ 
 ACT-S J013225$-$512903  &  $01:32:25$  &  $-51:29:03$  &  5.3  &  18.4  &  16.8$^{+3.6}_{-3.7}$  &  12.9  & ... \\ 
 ACT-S J013306$-$520003  &  $01:33:06$  &  $-52:00:03$  &  74.7  &  239.0  &  239.0$^{+3.2}_{-3.2}$  &  17.1  & ATG20 J013305$-$520003  \\ 
 ACT-S J013409$-$552612  &  $01:34:09$  &  $-55:26:12$  &  6.0  &  24.7  &  23.1$^{+4.2}_{-4.3}$  &  8.8  & ATG20 J013408$-$552616  \\ 
 ACT-S J013541$-$514943  &  $01:35:41$  &  $-51:49:43$  &  6.6  &  19.8  &  18.8$^{+3.0}_{-3.1}$  &  16.7  & ATG20 J013540$-$514945  \\ 
 ACT-S J013548$-$524414  &  $01:35:48$  &  $-52:44:14$  &  18.2  &  51.6  &  51.6$^{+2.8}_{-2.8}$  &  18.8  & ATG20 J013548$-$524417  \\ 
 ACT-S J013727$-$543942  &  $01:37:27$  &  $-54:39:42$  &  7.4  &  25.3  &  24.2$^{+3.5}_{-3.5}$  &  12.6  & ... \\ 
 ACT-S J013949$-$521739  &  $01:39:49$  &  $-52:17:39$  &  12.1  &  35.2  &  34.6$^{+2.9}_{-2.9}$  &  17.7  & ATG20 J013949$-$521746  \\ 
 ACT-S J014648$-$520232  &  $01:46:48$  &  $-52:02:32$  &  25.3  &  77.5  &  77.5$^{+3.1}_{-3.1}$  &  16.0  & ATG20 J014648$-$520233  \\ 
 ACT-S J015358$-$540649  &  $01:53:58$  &  $-54:06:49$  &  11.4  &  33.7  &  33.1$^{+3.0}_{-3.0}$  &  16.9  & ATG20 J015358$-$540653  \\ 
 ACT-S J015420$-$510750  &  $01:54:20$  &  $-51:07:50$  &  49.3  &  172.1  &  172.1$^{+3.5}_{-3.5}$  &  14.2  & ATG20 J015419$-$510751  \\ 
 ACT-S J015559$-$512538  &  $01:55:59$  &  $-51:25:38$  &  5.8  &  18.2  &  16.9$^{+3.2}_{-3.3}$  &  15.0  & ATG20 J015557$-$512545  \\ 
 ACT-S J015649$-$543940  &  $01:56:49$  &  $-54:39:40$  &  13.4  &  42.2  &  41.6$^{+3.2}_{-3.2}$  &  15.1  & ATG20 J015649$-$543949  \\ 
 ACT-S J015817$-$500415  &  $01:58:17$  &  $-50:04:15$  &  5.8  &  24.6  &  22.8$^{+4.4}_{-4.4}$  &  8.3  & ATG20 J015817$-$500419  \\ 
 ACT-S J015914$-$530902  &  $01:59:14$  &  $-53:09:02$  &  9.2  &  25.3  &  24.7$^{+2.8}_{-2.8}$  &  19.8  & ATG20 J015913$-$530853  \\ 
 ACT-S J020448$-$550257  &  $02:04:48$  &  $-55:02:57$  &  14.7  &  48.4  &  47.9$^{+3.2}_{-3.3}$  &  13.6  & ... \\ 
 ACT-S J020649$-$534528  &  $02:06:49$  &  $-53:45:28$  &  5.4  &  15.2  &  13.9$^{+2.9}_{-3.0}$  &  19.4  & ATG20 J020647$-$534543  \\ 
 ACT-S J020920$-$522921  &  $02:09:20$  &  $-52:29:21$  &  11.1  &  30.5  &  29.9$^{+2.8}_{-2.8}$  &  20.2  & ... \\ 
 ACT-S J021046$-$510100  &  $02:10:46$  &  $-51:01:00$  &  470.3  &  1677.6  &  1678.0$^{+3.6}_{-3.6}$  &  13.6  & ATG20 J021046$-$510101  \\ 
 ACT-S J021519$-$510435  &  $02:15:19$  &  $-51:04:35$  &  5.5  &  17.3  &  15.9$^{+3.3}_{-3.4}$  &  15.1  & ... \\ 
 ACT-S J021603$-$520007  &  $02:16:03$  &  $-52:00:07$  &  12.7  &  34.6  &  34.1$^{+2.7}_{-2.7}$  &  20.1  & ATG20 J021603$-$520012  \\ 
 ACT-S J021709$-$542750  &  $02:17:09$  &  $-54:27:50$  &  5.8  &  17.3  &  16.1$^{+3.1}_{-3.1}$  &  16.4  & ... \\ 
 ACT-S J021835$-$550354  &  $02:18:35$  &  $-55:03:54$  &  12.5  &  43.0  &  42.4$^{+3.5}_{-3.5}$  &  12.5  & ATG20 J021834$-$550350  \\ 
 ACT-S J022216$-$510627  &  $02:22:16$  &  $-51:06:27$  &  9.4  &  29.7  &  28.9$^{+3.2}_{-3.2}$  &  15.1  & ATG20 J022215$-$510629  \\ 
 ACT-S J022330$-$534737  &  $02:23:30$  &  $-53:47:37$  &  40.5  &  119.9  &  119.9$^{+3.0}_{-3.0}$  &  19.4  & ATG20 J022330$-$534740  \\ 
 ACT-S J022530$-$522547  &  $02:25:30$  &  $-52:25:47$  &  15.3  &  40.2  &  39.8$^{+2.6}_{-2.7}$  &  21.4  & ATG20 J022529$-$522555  \\ 
 ACT-S J022820$-$553732  &  $02:28:20$  &  $-55:37:32$  &  5.6  &  21.9  &  20.2$^{+4.1}_{-4.1}$  &  9.8  & ATG20 J022820$-$553725  \\ 
 ACT-S J022821$-$554601  &  $02:28:21$  &  $-55:46:01$  &  25.9  &  104.7  &  104.7$^{+4.0}_{-4.0}$  &  9.0  & ATG20 J022821$-$554603  \\ 
 ACT-S J022912$-$540324  &  $02:29:12$  &  $-54:03:24$  &  112.5  &  333.1  &  333.1$^{+3.0}_{-3.0}$  &  19.9  & ATG20 J022912$-$540324  \\ 
 ACT-S J022925$-$523226  &  $02:29:25$  &  $-52:32:26$  &  17.2  &  45.3  &  44.9$^{+2.6}_{-2.6}$  &  21.7  & ATG20 J022925$-$523226  \\ 
 ACT-S J023245$-$535634  &  $02:32:45$  &  $-53:56:34$  &  8.1  &  21.9  &  21.2$^{+2.7}_{-2.7}$  &  20.7  & ... \\ 
 ACT-S J023357$-$503014  &  $02:33:57$  &  $-50:30:14$  &  6.3  &  21.4  &  20.2$^{+3.5}_{-3.5}$  &  13.0  & ATG20 J023356$-$503020  \\ 
 ACT-S J023924$-$510824  &  $02:39:24$  &  $-51:08:24$  &  8.5  &  25.2  &  24.4$^{+3.0}_{-3.0}$  &  16.8  & ... \\ 
 ACT-S J024040$-$542942  &  $02:40:40$  &  $-54:29:42$  &  6.9  &  21.7  &  20.7$^{+3.2}_{-3.2}$  &  15.3  & ATG20 J024040$-$542933  \\ 
 ACT-S J024154$-$534541  &  $02:41:54$  &  $-53:45:41$  &  6.7  &  17.8  &  16.9$^{+2.7}_{-2.8}$  &  20.9  & ... \\ 
 ACT-S J024313$-$510512  &  $02:43:13$  &  $-51:05:12$  &  11.5  &  35.1  &  34.5$^{+3.1}_{-3.1}$  &  16.2  & ATG20 J024313$-$510517  \\ 
 ACT-S J024539$-$525756  &  $02:45:39$  &  $-52:57:56$  &  7.6  &  19.1  &  18.4$^{+2.6}_{-2.6}$  &  23.4  & ... \\ 
 ACT-S J024614$-$495346  &  $02:46:14$  &  $-49:53:46$  &  10.3  &  42.8  &  41.9$^{+4.2}_{-4.2}$  &  8.5  & ATG20 J024614$-$495350  \\ 
 ACT-S J024947$-$555619  &  $02:49:47$  &  $-55:56:19$  &  5.4  &  22.8  &  20.9$^{+4.4}_{-4.6}$  &  8.2  & ATG20 J024948$-$555615  \\ 
 ACT-S J025112$-$520822  &  $02:51:12$  &  $-52:08:22$  &  5.7  &  15.0  &  13.9$^{+2.7}_{-2.8}$  &  21.8  & ... \\ 
 ACT-S J025204$-$514554  &  $02:52:04$  &  $-51:45:54$  &  5.4  &  14.4  &  13.2$^{+2.8}_{-2.8}$  &  20.7  & ... \\ 
 ACT-S J025328$-$544151  &  $02:53:28$  &  $-54:41:51$  &  246.7  &  837.5  &  837.5$^{+3.4}_{-3.4}$  &  15.3  & ATG20 J025329$-$544151  \\ 
 ACT-S J025629$-$522711  &  $02:56:29$  &  $-52:27:11$  &  5.3  &  14.1  &  12.8$^{+2.8}_{-2.9}$  &  21.0  & ... \\ 
 ACT-S J025838$-$505200  &  $02:58:38$  &  $-50:52:00$  &  33.2  &  113.6  &  113.6$^{+3.4}_{-3.4}$  &  14.6  & ATG20 J025838$-$505204  \\ 
 ACT-S J025849$-$533210  &  $02:58:49$  &  $-53:32:10$  &  7.2  &  19.5  &  18.6$^{+2.7}_{-2.8}$  &  20.6  & ... \\ 
 ACT-S J030056$-$510229  &  $03:00:56$  &  $-51:02:29$  &  11.2  &  36.8  &  36.1$^{+3.3}_{-3.3}$  &  14.0  & ATG20 J030055$-$510229  \\ 
 ACT-S J030132$-$525603  &  $03:01:32$  &  $-52:56:03$  &  15.0  &  39.7  &  39.3$^{+2.6}_{-2.7}$  &  21.6  & ... \\ 
 ACT-S J030327$-$523427  &  $03:03:27$  &  $-52:34:27$  &  24.8  &  66.4  &  66.4$^{+2.7}_{-2.7}$  &  21.0  & ATG20 J030328$-$523433  \\ 
 ACT-S J030616$-$552808  &  $03:06:16$  &  $-55:28:08$  &  15.2  &  56.4  &  56.5$^{+3.7}_{-3.7}$  &  10.9  & ATG20 J030616$-$552808  \\ 
 ACT-S J031206$-$554135  &  $03:12:06$  &  $-55:41:35$  &  18.3  &  72.7  &  72.7$^{+4.0}_{-4.0}$  &  9.5  & ATG20 J031207$-$554133  \\ 
 ACT-S J031426$-$510430  &  $03:14:26$  &  $-51:04:30$  &  29.0  &  90.6  &  90.5$^{+3.1}_{-3.1}$  &  15.2  & ATG20 J031425$-$510431  \\ 
 ACT-S J031823$-$533148  &  $03:18:23$  &  $-53:31:48$  &  18.6  &  51.6  &  51.6$^{+2.8}_{-2.8}$  &  19.6  & ... \\ 
 ACT-S J031910$-$500031  &  $03:19:10$  &  $-50:00:31$  &  7.5  &  30.2  &  28.9$^{+4.1}_{-4.1}$  &  9.3  & ... \\ 
 ACT-S J032207$-$535419  &  $03:22:07$  &  $-53:54:19$  &  6.5  &  18.3  &  17.3$^{+2.9}_{-2.9}$  &  19.6  & ... \\ 
 ACT-S J032212$-$504229  &  $03:22:12$  &  $-50:42:29$  &  11.0  &  35.5  &  34.8$^{+3.2}_{-3.3}$  &  14.5  & ... \\ 
 ACT-S J032327$-$522627  &  $03:23:27$  &  $-52:26:27$  &  14.9  &  41.0  &  40.5$^{+2.8}_{-2.8}$  &  19.7  & ATG20 J032327$-$522630  \\ 
 ACT-S J032650$-$533658  &  $03:26:50$  &  $-53:36:58$  &  17.0  &  48.0  &  47.6$^{+2.8}_{-2.8}$  &  18.9  & ATG20 J032650$-$533701  \\ 
 ACT-S J033002$-$503518  &  $03:30:02$  &  $-50:35:18$  &  6.5  &  21.9  &  20.7$^{+3.4}_{-3.4}$  &  13.5  & ATG20 J033002$-$503519  \\ 
 ACT-S J033114$-$524149  &  $03:31:14$  &  $-52:41:49$  &  9.0  &  24.4  &  23.7$^{+2.7}_{-2.7}$  &  21.1  & ATG20 J033114$-$524148  \\ 
 ACT-S J033126$-$525829  &  $03:31:26$  &  $-52:58:29$  &  11.9  &  31.6  &  31.0$^{+2.7}_{-2.7}$  &  21.5  & ATG20 J033126$-$525830  \\ 
 ACT-S J033133$-$515352  &  $03:31:33$  &  $-51:53:52$  &  5.9  &  16.1  &  15.0$^{+2.8}_{-2.9}$  &  19.8  & ... \\ 
 ACT-S J033444$-$521851  &  $03:34:44$  &  $-52:18:51$  &  7.1  &  19.2  &  18.4$^{+2.8}_{-2.8}$  &  20.4  & ... \\ 
 ACT-S J033554$-$543028  &  $03:35:54$  &  $-54:30:28$  &  14.8  &  44.0  &  43.6$^{+3.0}_{-3.0}$  &  16.6  & ATG20 J033553$-$543025  \\ 
 ACT-S J034157$-$515140  &  $03:41:57$  &  $-51:51:40$  &  6.3  &  16.6  &  15.7$^{+2.7}_{-2.7}$  &  22.1  & ... \\ 
 ACT-S J034348$-$524112  &  $03:43:48$  &  $-52:41:12$  &  8.3  &  21.9  &  21.2$^{+2.7}_{-2.7}$  &  21.5  & ATG20 J034349$-$524116  \\ 
 ACT-S J034940$-$540111  &  $03:49:40$  &  $-54:01:11$  &  7.4  &  21.7  &  20.7$^{+3.0}_{-3.0}$  &  17.1  & ATG20 J034941$-$540106  \\ 
 ACT-S J035128$-$514254  &  $03:51:28$  &  $-51:42:54$  &  50.7  &  145.2  &  145.2$^{+2.9}_{-2.9}$  &  21.6  & ATG20 J035128$-$514254  \\ 
 ACT-S J035343$-$534553  &  $03:53:43$  &  $-53:45:53$  &  5.6  &  16.3  &  15.0$^{+3.0}_{-3.1}$  &  17.9  & ... \\ 
 ACT-S J035700$-$495549  &  $03:57:00$  &  $-49:55:49$  &  14.1  &  58.3  &  58.3$^{+4.1}_{-4.1}$  &  8.7  & ATG20 J035700$-$495547  \\ 
 ACT-S J035840$-$543403  &  $03:58:40$  &  $-54:34:03$  &  8.3  &  27.4  &  26.4$^{+3.3}_{-3.4}$  &  14.0  & ... \\ 
 ACT-S J040401$-$552022  &  $04:04:01$  &  $-55:20:22$  &  7.0  &  27.6  &  26.3$^{+4.0}_{-4.0}$  &  10.2  & ATG20 J040400$-$552023  \\ 
 ACT-S J040621$-$503509  &  $04:06:21$  &  $-50:35:09$  &  5.5  &  20.5  &  18.8$^{+3.9}_{-4.0}$  &  10.8  & ... \\ 
 ACT-S J041137$-$514918  &  $04:11:37$  &  $-51:49:18$  &  22.8  &  72.5  &  72.5$^{+3.2}_{-3.2}$  &  14.9  & ATG20 J041137$-$514923  \\ 
 ACT-S J041247$-$560044  &  $04:12:47$  &  $-56:00:44$  &  5.4  &  23.4  &  21.4$^{+4.5}_{-4.6}$  &  8.1  & ATG20 J041247$-$560035  \\ 
 ACT-S J041313$-$533157  &  $04:13:13$  &  $-53:31:57$  &  13.2  &  39.5  &  38.9$^{+3.0}_{-3.0}$  &  16.6  & ATG20 J041313$-$533200  \\ 
 ACT-S J041959$-$545618  &  $04:19:59$  &  $-54:56:18$  &  7.2  &  23.8  &  22.7$^{+3.4}_{-3.4}$  &  13.8  & ... \\ 
 ACT-S J042503$-$533201  &  $04:25:03$  &  $-53:32:01$  &  49.8  &  152.1  &  152.1$^{+3.1}_{-3.1}$  &  18.6  & ATG20 J042504$-$533158  \\ 
 ACT-S J042842$-$500532  &  $04:28:42$  &  $-50:05:32$  &  33.1  &  154.7  &  154.7$^{+4.7}_{-4.7}$  &  6.9  & ATG20 J042842$-$500534  \\ 
 ACT-S J042852$-$543001  &  $04:28:52$  &  $-54:30:01$  &  6.8  &  21.8  &  20.8$^{+3.3}_{-3.3}$  &  14.5  & ATG20 J042852$-$543007  \\ 
 ACT-S J042906$-$534943  &  $04:29:06$  &  $-53:49:43$  &  31.7  &  87.4  &  87.4$^{+2.8}_{-2.8}$  &  19.9  & ATG20 J042908$-$534940  \\ 
 ACT-S J043221$-$510926  &  $04:32:21$  &  $-51:09:26$  &  26.5  &  86.8  &  86.8$^{+3.3}_{-3.3}$  &  14.1  & ATG20 J043221$-$510925  \\ 
 ACT-S J043651$-$521632  &  $04:36:51$  &  $-52:16:32$  &  14.1  &  38.4  &  38.0$^{+2.7}_{-2.7}$  &  20.4  & ATG20 J043652$-$521639  \\ 
 ACT-S J044115$-$543848  &  $04:41:15$  &  $-54:38:48$  &  6.1  &  19.4  &  18.2$^{+3.3}_{-3.3}$  &  14.9  & ... \\ 
 ACT-S J044158$-$515456  &  $04:41:58$  &  $-51:54:56$  &  30.4  &  87.5  &  87.5$^{+2.9}_{-2.9}$  &  18.2  & ATG20 J044158$-$515453  \\ 
 ACT-S J044502$-$523426  &  $04:45:02$  &  $-52:34:26$  &  9.1  &  25.5  &  24.8$^{+2.8}_{-2.8}$  &  19.4  & ... \\ 
 ACT-S J044746$-$515103  &  $04:47:46$  &  $-51:51:03$  &  9.7  &  27.2  &  26.5$^{+2.8}_{-2.8}$  &  19.3  & ATG20 J044748$-$515100  \\ 
 ACT-S J044821$-$504140  &  $04:48:21$  &  $-50:41:40$  &  10.8  &  37.1  &  36.3$^{+3.5}_{-3.4}$  &  13.1  & ATG20 J044822$-$504133  \\ 
 ACT-S J045028$-$534659  &  $04:50:28$  &  $-53:46:59$  &  8.9  &  24.8  &  24.1$^{+2.8}_{-2.8}$  &  19.5  & ... \\ 
 ACT-S J045103$-$493632  &  $04:51:03$  &  $-49:36:32$  &  12.1  &  58.2  &  58.2$^{+4.8}_{-4.8}$  &  6.3  & ATG20 J045102$-$493626  \\ 
 ACT-S J045239$-$530637  &  $04:52:39$  &  $-53:06:37$  &  8.4  &  23.3  &  22.6$^{+2.8}_{-2.8}$  &  19.3  & ATG20 J045238$-$530635  \\ 
 ACT-S J045503$-$553119  &  $04:55:03$  &  $-55:31:19$  &  9.8  &  36.7  &  35.8$^{+3.8}_{-3.8}$  &  10.8  & ATG20 J045503$-$553112  \\ 
 ACT-S J045559$-$530236  &  $04:55:59$  &  $-53:02:36$  &  6.9  &  18.3  &  17.4$^{+2.7}_{-2.7}$  &  21.2  & ATG20 J045558$-$530239  \\ 
 ACT-S J050018$-$532122  &  $05:00:18$  &  $-53:21:22$  &  10.6  &  28.2  &  27.6$^{+2.7}_{-2.7}$  &  21.3  & ATG20 J050019$-$532121  \\ 
 ACT-S J050401$-$502314  &  $05:04:01$  &  $-50:23:14$  &  15.8  &  59.6  &  59.5$^{+3.8}_{-3.8}$  &  10.7  & ATG20 J050401$-$502313  \\ 
 ACT-S J051355$-$505543  &  $05:13:55$  &  $-50:55:43$  &  6.5  &  22.0  &  20.9$^{+3.4}_{-3.5}$  &  13.4  & ATG20 J051355$-$505541  \\ 
 ACT-S J051812$-$514359  &  $05:18:12$  &  $-51:43:59$  &  8.7  &  24.7  &  23.9$^{+2.9}_{-2.9}$  &  18.6  & ATG20 J051811$-$514404  \\ 
 ACT-S J052044$-$550832  &  $05:20:44$  &  $-55:08:32$  &  7.0  &  24.1  &  22.9$^{+3.5}_{-3.5}$  &  12.7  & ATG20 J052045$-$550824  \\ 
 ACT-S J052317$-$530836  &  $05:23:17$  &  $-53:08:36$  &  8.4  &  22.9  &  22.1$^{+2.7}_{-2.7}$  &  20.5  & ATG20 J052318$-$530837  \\ 
 ACT-S J052743$-$542614  &  $05:27:43$  &  $-54:26:14$  &  8.9  &  26.0  &  25.3$^{+2.9}_{-2.9}$  &  17.3  & ATG20 J052743$-$542616  \\ 
 ACT-S J053117$-$550431  &  $05:31:17$  &  $-55:04:31$  &  14.1  &  47.9  &  47.3$^{+3.3}_{-3.4}$  &  13.0  & ... \\ 
 ACT-S J053208$-$531035  &  $05:32:08$  &  $-53:10:35$  &  18.9  &  51.2  &  51.2$^{+2.7}_{-2.7}$  &  20.3  & ATG20 J053208$-$531035  \\ 
 ACT-S J053323$-$554941  &  $05:33:23$  &  $-55:49:41$  &  16.7  &  65.7  &  65.7$^{+3.9}_{-3.9}$  &  9.7  & ATG20 J053324$-$554936  \\ 
 ACT-S J053458$-$543903  &  $05:34:58$  &  $-54:39:03$  &  10.0  &  30.8  &  30.1$^{+3.1}_{-3.1}$  &  15.7  & ATG20 J053458$-$543901  \\ 
 ACT-S J053909$-$551059  &  $05:39:09$  &  $-55:10:59$  &  8.9  &  30.6  &  29.7$^{+3.5}_{-3.5}$  &  12.8  & ATG20 J053909$-$551059  \\ 
 ACT-S J054025$-$530354  &  $05:40:25$  &  $-53:03:54$  &  11.5  &  31.7  &  31.1$^{+2.8}_{-2.8}$  &  19.9  & ATG20 J054025$-$530346  \\ 
 ACT-S J054029$-$535628  &  $05:40:29$  &  $-53:56:28$  &  5.5  &  14.8  &  13.6$^{+2.8}_{-2.8}$  &  21.2  & ATG20 J054029$-$535632  \\ 
 ACT-S J054046$-$541825  &  $05:40:46$  &  $-54:18:25$  &  133.4  &  416.2  &  416.2$^{+3.1}_{-3.1}$  &  17.4  & ATG20 J054045$-$541821  \\ 
 ACT-S J054223$-$514259  &  $05:42:23$  &  $-51:42:59$  &  27.4  &  79.7  &  79.7$^{+2.9}_{-2.9}$  &  18.1  & ATG20 J054223$-$514257  \\ 
 ACT-S J054357$-$553206  &  $05:43:57$  &  $-55:32:06$  &  6.2  &  22.2  &  20.9$^{+3.7}_{-3.7}$  &  11.3  & ... \\ 
 ACT-S J054830$-$521836  &  $05:48:30$  &  $-52:18:36$  &  9.3  &  25.2  &  24.6$^{+2.7}_{-2.7}$  &  20.2  & ... \\ 
 ACT-S J054943$-$524629  &  $05:49:43$  &  $-52:46:29$  &  63.4  &  180.3  &  180.3$^{+2.8}_{-2.8}$  &  21.8  & ATG20 J054943$-$524625  \\ 
 ACT-S J055047$-$530502  &  $05:50:47$  &  $-53:05:02$  &  9.0  &  24.5  &  23.8$^{+2.8}_{-2.8}$  &  20.1  & ... \\ 
 ACT-S J055152$-$552642  &  $05:51:52$  &  $-55:26:42$  &  8.6  &  31.5  &  30.4$^{+3.7}_{-3.7}$  &  11.1  & ATG20 J055152$-$552632  \\ 
 ACT-S J055811$-$502957  &  $05:58:11$  &  $-50:29:57$  &  16.0  &  57.5  &  57.5$^{+3.6}_{-3.6}$  &  11.8  & ATG20 J055811$-$502948  \\ 
 ACT-S J055830$-$532640  &  $05:58:30$  &  $-53:26:40$  &  6.1  &  16.6  &  15.6$^{+2.8}_{-2.8}$  &  19.7  & ATG20 J055830$-$532631  \\ 
 ACT-S J055946$-$502656  &  $05:59:46$  &  $-50:26:56$  &  12.5  &  45.9  &  45.2$^{+3.7}_{-3.7}$  &  11.2  & ATG20 J055947$-$502652  \\ 
 ACT-S J060213$-$542509  &  $06:02:13$  &  $-54:25:09$  &  12.5  &  38.3  &  37.7$^{+3.1}_{-3.1}$  &  15.9  & ATG20 J060212$-$542507  \\ 
 ACT-S J060749$-$525747  &  $06:07:49$  &  $-52:57:47$  &  13.1  &  36.0  &  35.5$^{+2.8}_{-2.7}$  &  19.6  & ATG20 J060749$-$525744  \\ 
 ACT-S J060849$-$545650  &  $06:08:49$  &  $-54:56:50$  &  38.4  &  136.9  &  136.9$^{+3.6}_{-3.6}$  &  13.7  & ATG20 J060849$-$545642  \\ 
 ACT-S J061715$-$530615  &  $06:17:15$  &  $-53:06:15$  &  13.1  &  39.6  &  39.0$^{+3.0}_{-3.1}$  &  16.2  & ... \\ 
 ACT-S J061846$-$532948  &  $06:18:46$  &  $-53:29:48$  &  6.3  &  18.7  &  17.6$^{+3.0}_{-3.1}$  &  16.8  & ... \\ 
 ACT-S J061955$-$542718  &  $06:19:55$  &  $-54:27:18$  &  12.1  &  39.1  &  38.5$^{+3.3}_{-3.3}$  &  13.8  & ATG20 J061955$-$542713  \\ 
 ACT-S J062142$-$524136  &  $06:21:42$  &  $-52:41:36$  &  37.3  &  118.9  &  118.9$^{+3.2}_{-3.2}$  &  17.4  & ATG20 J062143$-$524132  \\ 
 ACT-S J062552$-$543856  &  $06:25:52$  &  $-54:38:56$  &  28.5  &  105.9  &  105.9$^{+3.7}_{-3.7}$  &  10.8  & ATG20 J062552$-$543850  \\ 
 ACT-S J062620$-$534136  &  $06:26:20$  &  $-53:41:36$  &  15.1  &  46.7  &  46.3$^{+3.1}_{-3.1}$  &  15.8  & ATG20 J062620$-$534151  \\ 
 ACT-S J062649$-$543233  &  $06:26:49$  &  $-54:32:33$  &  8.9  &  32.8  &  31.8$^{+3.7}_{-3.7}$  &  11.0  & ATG20 J062648$-$543214  \\ 
 ACT-S J063200$-$540501  &  $06:32:00$  &  $-54:05:01$  &  6.1  &  21.0  &  19.7$^{+3.5}_{-3.6}$  &  12.7  & ATG20 J063201$-$540455  \\ 
 ACT-S J064111$-$520223  &  $06:41:11$  &  $-52:02:23$  &  5.8  &  19.4  &  18.1$^{+3.4}_{-3.5}$  &  13.5  & ... \\ 
 ACT-S J064319$-$535850  &  $06:43:19$  &  $-53:58:50$  &  32.1  &  104.5  &  104.5$^{+3.3}_{-3.3}$  &  14.0  & ATG20 J064320$-$535846  \\ 
 ACT-S J064629$-$545120  &  $06:46:29$  &  $-54:51:20$  &  9.7  &  37.7  &  36.7$^{+3.9}_{-3.9}$  &  10.1  & ATG20 J064629$-$545116  \\ 

\enddata

\tablenotetext{a}{Flux density as measured directly from the ACT \arone\  map.} 
\tablenotetext{b}{Deboosted flux densities as described in Section \ref{subsec:deboosting} for sources with $S_{\rm m} < 50$~mJy. For sources measured flux in above 50 mJy, the measured flux together with S/N derived errors are reported (See Section \ref{subsec:deboosting}.). }
\tablenotetext{b}{Integration time per square arcminute}
\label{tab:sources}
\end{deluxetable}

\end{document}